\documentclass[fleqn]{Configurations/2023SCGE}
\usepackage{graphicx}
\usepackage{mathrsfs}
\usepackage{placeins} 
\usepackage{float}
\usepackage{stfloats}
\usepackage{tabularx}
\usepackage[numbers,sort&compress]{natbib}
\usepackage{lscape}
\usepackage{amsmath,amssymb}
\usepackage{booktabs}
\usepackage{array,makecell}
\usepackage{diagbox}
\usepackage{ragged2e}
\usepackage{url}
\usepackage{rotating}
\usepackage{multirow}   
\usepackage[normalem]{ulem}
\usepackage{amsmath}
\newcolumntype{C}[1]{>{\centering\arraybackslash}p{#1}}

\newcolumntype{Z}{>{\Centering\arraybackslash}X}

\begin{document}

\raggedcolumns

\ensubject{subject}
\ArticleType{Article}
\SpecialTopic{SPECIAL TOPIC: }
\Year{2023} \Month{January} \Vol{66} \No{1} \DOI{??} \ArtNo{000000} \ReceiveDate{March 30, 2026}
\AcceptDate{??}

\title{Growth from KiDS: Stellar mass assembly of galaxies since z=2 in the light of Kilo-Degree Survey}
{KiDS DR4 GSMF}

\AuthorCitation{Wang Q, Katsianis A, Napolitano N R, et al.}

\author[1]{Qingshan Wang}{}
\author[1]{Antonios Katsianis}{katsianis@sysu.edu.cn}
\author[2,3]{Nicola R. Napolitano}{}
\author[1]{Linghua Xie}{}
\author[4]{Rui Li}{}
\author[5,6,7]{Xiaohu Yang}{}
\author[1]{\\Baitian Tang}{}
\author[5,6]{Jiaxin Han}{}
\author[5,6]{Zhenlin Tan}{}
\author[8]{Lizhi Xie}{}
\author[1]{Ying Tang}{}
\author[1]{Yuchang Li}{}
\author[1]{Hangxin Pu}{}

\address[1]{School of Physics and Astronomy, Sun Yat-sen University, Zhuhai Campus, 2 Daxue Road, Xiangzhou District, Zhuhai, 519082, China}
\address[2]{Department of Physics E. Pancini, University Federico II, Via Cinthia 21, I-80126 Naples, Italy}
\address[3]{INAF – Osservatorio Astronomico di Capodimonte, Salita Moiariello 16, I-80131 Napoli, Italy}
\address[4]{Institute for Astrophysics, School of Physics, Zhengzhou University, Zhengzhou, 450001, China}
\address[5]{Department of Astronomy, Shanghai Jiao Tong University, Shanghai 200240, China}
\address[6]{State Key Laboratory of Dark Matter Physics, Key Laboratory for Particle Astrophysics and Cosmology (MOE), \\ \& Shanghai Key Laboratory for Particle Physics and Cosmology, Shanghai Jiao Tong University, Shanghai 200240, China}
\address[7]{Tsung-Dao Lee Institute and State Key Laboratory of Dark Matter Physics, Shanghai Jiao Tong University, Shanghai 202110, China}
\address[8]{Astrophysics Center, Tianjin Normal University, Tianjin 300387, China}

\abstract{We measure the galaxy stellar mass function (GSMF) at $z = 0-2$ using 28M galaxies from the KiDS DR4 which spans an effective survey area of 676.9 deg$^2$. Combining nine-band photometry ($u$--$K_s$) and redshifts derived using deep learning photometry, we use five SED fitting configurations, CIGALE (combined with delayed exponential declining star formation history) / LePhare (combined with exponential declining star formation history) and CB19/BC03/M05 Stellar Population Synthesis (SPS). We present the impact of the systematics related to the code, SPS modeling and Star formation History (SFH) , which combined can introduce systematics of $\sim 0.6$ dex. The GSMFs outlined in this work  agree well with previous studies (within the uncertainties introduced by systematics) that typically are able to cover measurements for the GSMF only up to $\log(M_*/M_\odot)\sim 11.75$, due to their limited volume. The large KiDS area and galaxy counts, allows us to gain some  insights on the ``extreme" high-mass end, and our GSMFs imply that possibly there are more massive galaxies than we would expect, from a simple exponential cut-off, at $\log(M_*/M_\odot)>11.75$. However, we heed caution that despite advances done both in redshift estimation techniques and SPS, future studies are required to confirm this behavior. We  consider the effects of  Eddington-bias (EB) to our observations via two different methods and compare our results with the predictions of semi-analytic models and hydrodynamical simulations. We use the Mean Absolute Difference (MAD) of residuals as a diagnostic metric. Due to inconsistencies ($> 0.5$ dex) between observations and Eddington biased simulations found, especially at the high mass end, we suggest that some theoretical models would achieve better performance once parameter tuning takes into account the effects of EB prior to the tuning of the uncertain parameters involved related to stellar physics and feedback.}

\maketitle

\begin{multicols}{2}
    \section{Introduction}
    \label{sec: intro}

    The Galaxy Stellar Mass Function (GSMF) is defined as the comoving number density of galaxies per unit stellar mass interval. By tracing the abundance of galaxies as a function of stellar mass at different epochs, the GSMF encapsulates the cumulative outcome of star formation, galaxy assembly, and feedback processes, and therefore serves as a fundamental constraint on theoretical models of galaxy formation and evolution \citep[e.g.][]{White1978,WhiteFrenk1991,Somerville2015,WechslerTinker2018,Katsianis2023}. 

    Owing to its importance, numerous observational studies have explored the evolution of the GSMF \citep{Gonzalez2011,Baldry2012,Wang2024}, shedding light on the cosmic build-up of stellar mass, the fraction of mass locked in stars, and the growth and quenching of massive galaxies. For example, studies of the GSMF have revealed that the number density of star-forming galaxies increases by several orders of magnitude between $z\sim7.5$ and $z\sim2$, marking the era of rapid stellar mass assembly, after which massive galaxies become predominantly quiescent—a phenomenon known as \emph{downsizing} \citep{Weaver2023}. Some additional insights come  from intermediate redshifts ($z\sim2.5$), where the GSMF shows an excess of massive, red systems something that implies that massive galaxies have been growing and experiencing quenching fast (within 2-3 billion years). Studies of the GSMF at low redshift ($z<0.1$), have also revealed that  only $\sim4.9\%$ of the baryonic mass is locked into stars within galaxies \citep{Driver2022}, and the growth of massive red galaxies continues at $z<1$ mainly through mergers \citep{Beare2019}.

    Due to its importance there have been efforts to understand and replicate the GSMF of galaxies using theoretical or computational methods \citep{Katsianis2025}. Consequently, the GSMF has become one of the key benchmarks for testing and calibrating modern models of galaxy formation. On the one hand, semi-analytic models (SAMs) leverage dark matter merger trees from N-body simulations, populating them with galaxies using physical prescriptions for key processes such as gas cooling, star formation, and feedback from both stars and active galactic nuclei (AGN) \citep{Benson2003,Croton2006,Bower2006,Somerville2008,Henriques2015,WhiteFrenk1991,Kauffmann1993,Somerville2015,Lagos2018,Fontanot2025}. These models can efficiently explore large parameter spaces and can predict GSMF evolution across cosmic time. On the other hand, hydrodynamical simulations, such as IllustrisTNG \citep{Pillepich2018,Nelson2019, Vogelsberger2020, Li2025, Ramesh2025}, EAGLE \citep{Schaye2015,Crain2015,Katsianis2017,Baes2020,Qiao2024,Das2024,Proctor2025} and Simba \citep{Dave2019,Katsianis2021,Yang2024,Thomas2025}, solve the coupled evolution of gas and dark matter, incorporating complex sub-grid physics for star formation and feedback. While they provide more detailed predictions for galaxy properties, their results are often affected by limited resolution and cosmological volume \citep{Zhao2020}. SAMs and simulations currently are supposed to be able to reproduce broadly the observed GSMF across most  mass ranges ($10^6$–$10^{12}\,M_\odot$) and redshifts, however this is done by using completely different Philosophies/methodologies/feedback prescriptions, and extensive tuning against the observed GSMFs, something that highlights some gaps in our understanding of key physical processes \citep{Somerville2015,Naab2017,Crain2023,Katsianis2025}, especially for extremely high mass objects. Moreover, such comparisons typically neglect observational systematics and effects like Eddington bias, which can lead to misinterpretations of a model's true performance.

    The determination of the observed GSMF critically depends on survey characteristics, particularly the balance between area and depth. Wide surveys such as SDSS provide excellent statistics for massive galaxies and yield precise constraints at the high-mass end \citep{Bernardi2010,Moustakas2013}, but lack the depth required to probe higher redshifts. Conversely, deep pencil-beam surveys like COSMOS and GOODS can reach low-mass galaxies at high redshift, yet their limited volume leads to significant cosmic variance \citep{Ilbert2010,Santini2012,Grazian2015}. Recent large-area, moderately deep surveys have begun to bridge these regimes by combining substantial sky coverage with sufficient depth. Within this context, \citet{Xielh2023} used the fourth data release of the Kilo-Degree Survey (KiDS DR4, conducted by the VST \citep{Capaccioli2005}) to demonstrate the power of KiDS+VIKING nine-band photometry ($u,g,r,i,Z,Y,J,H,K_s$) for deriving robust stellar mass and SFR estimates with CIGALE and LePhare, providing a reliable stellar mass  catalog for galaxies with spectroscopic redshifts up to $z \sim 0.9$. These estimates of the GSMF from \citet{Xielh2023} were presented up to $z \sim 0.4$ with the limitation though that the stellar mass histograms were just normalized to previous literature data (i.e. there was not an analysis of the volume involved, or a dedicated procedure using a selection function).

    Following the SED fitting framework of \citet{Xielh2023}, we use  KiDS DR4 to perform this time a comprehensive GSMF analysis covering an extended redshift range of $0.001 < z < 2.0$. We update the catalogue and this deeper, expanded optical--NIR dataset covers around 676.9 deg$^2$ (after masking, 2 times of \citet{Wright2019}, who used DR3 data) down to $r \sim 25$ mag, enabling more robust GSMF measurements across a wider stellar mass range while mitigating cosmic variance, and is particularly advantageous for constraining the rare, massive end of the GSMF \citep{Driver2022,Wright2024}. Photometric redshifts are derived using morphology-assisted photometric redshifts (``morphotometric'' redshifts) with the deep-learning tool GaZNet \citep[GaZNet-z hereafter;][]{Li2022}. Stellar masses are obtained through spectral energy distribution (SED) fitting with CIGALE \citep{Boquien2019} and LePhare \citep{Arnouts1999,Ilbert2006}, adopting widely used stellar population synthesis models, BC03 \citep{Bruzual2003},  M05 \citep{Maraston2005} and CB19 \citep{CB19}, to cover a wide range of codes/template/star formation histories, make sensible comparisons with previous studies,  and take into account the effect of the systematics coming from different stellar population analysis set-ups (see Xie et al. 2023\citep{Xielh2023} for a discussion).

    The paper is organized as follows. Section~\ref{sec:sample_selection} describes the KiDS DR4 dataset, sample construction, and our analysis methodology, including photometric redshift estimation and stellar mass derivation. Section~\ref{sec:GSMF-methods} presents our GSMF estimation using the $1/V_{\rm max}$ method. Section~\ref{sec:GSMF-results-discussion} shows the results compared with previous observational studies and discusses comparisons with theoretical models including semi-analytical models and hydrodynamical simulations. Section~\ref{sec:summary} provides our summary and conclusions. Throughout this work, we assume a flat $\Lambda$CDM cosmology with $H_0 = 70$ km s$^{-1}$ Mpc$^{-1}$, $\Omega_m = 0.3$, and $\Omega_\Lambda = 0.7$.
    \section{Sample construction, selection and physical properties estimation from KiDS DR4}
    \label{sec:sample_selection}

    In this section, we describe our sample selection and construction from KiDS DR4. We also discuss the photometric redshift estimation, stellar mass estimation and the stellar mass completeness of the sample, which serve as the basis for the GSMF estimation in the next section.
    \subsection{Sample Construction}
    \label{sec:sample_construction}

    We start with the total KiDS DR4 catalog, which contains over 100 million sources \citep{Kuijken2019}.  In our work, we go beyond using the cross matched GAMA KiDS DR4 that was used in \citet{Xielh2023} and the KV450 used in \citet{Wright2019}. We follow a multi-step process to select galaxies from the total available sources, which is described in the following sections. 
    
    \subsubsection{From total sources to base sample}
    \label{sec:total_to_base}

    We first apply basic quality cuts to all sources in the KiDS DR4 catalog to obtain a base sample. The selections include:
    \begin{itemize}
        \item \textit{9-band MASK:} This parameter indicates the availability and quality of the data at a given sky position \cite{Kuijken2019}. We selected $\text{MASK} \le 1$ to exclude defects in each band image, which were automatically detected by the photometric pipeline or manually flagged. This procedure mitigates the contamination from scattered and diffracted light (e.g., stellar artifacts). We include an additional tile exclusion that if the ratio of this MASK selection in the tile is larger than 95\%, we abandon this tile.
        \item \textit{Asteroid:} Like \citet{Kuijken2015}, we exclude asteroids by excluding extreme color $(g-r \le 1.5) | (i-r \le 1.5) $.
        \item \textit{IMAFLAGS\_ISO:} This parameter indicates the quality of the photometric isophotes. As described in the first and second KiDS data release \citep{Jong2015}, by selecting \texttt{IMAFLAGS\_ISO = 0}, we exclude areas with spikes, saturated core reflection halos and bad pixels and choose only the highest-quality sources.
    \end{itemize}

    After these steps we get the an effective survey area which is 676.9 $deg^2$.
    \subsubsection{From a base sample to a 9-band sample}
    \label{sec:base_to_9band}

    We then select sources from the base sample that possess complete 9-band photometric data, which are essential for reliable SED fitting. We do this by applying the following  magnitude constraints to ensure we include all potential galaxies while minimizing artifacts following \citet{Xielh2023}:
    \begin{itemize}
        \item \textit{MAG\_GAAP\_x:}\footnote{MAG\_GAAP\_x denotes the aperture magnitude measured in the x-band using the Gaussian Aperture and PSF (GAAP) pipeline, defined as the flux weighted by a consistent elliptical Gaussian aperture on a PSF-homogenized image to ensure robust, seeing-independent color measurements across all bands\citep{Kuijken2019}.} We require sources to satisfy $0 < \text{MAG\_GAAP\_x} < 44$. This broad constraint ensures we have sufficient photometric coverage to perform accurate SED fitting and derive reliable stellar mass estimates.
        \item \textit{MAG\_AUTO:}\footnote{ MAG\_AUTO denotes the quasi-total magnitude of the r-band estimated by the SExtractor (Bertin \& Arnouts 1996)\citep{Kuijken2019}.} We also apply $0 < \text{MAG\_AUTO} < 44$ to guaranty a measurement.
        \item \textit{FLAG\_GAAP\_x:}\footnote{FLAG\_GAAP\_x denotes the successness of GAAP magnitude measurement, most likely overlapped with MAG\_GAAP\_x selection.} We only select sources with pristine measurements, requiring the sum of all flags to be zero ($\sum \text{FLAG\_GAAP\_x} = 0$), to ensure reliable data across all photometric bands.
    \end{itemize}

    We expect that this step will introduce a selection effect and to correct for this we apply the Inverse Probability Selection Weighting (IPSW) method \citep{Horvitz1952} in our GSMF estimation, which will be described in more detail in Section~\ref{sec:IPSW}.
    \subsubsection{From a 9-band sample to a pure galaxy sample}
    \label{sec:classification}

    In order to build a pure\footnote{Including only galaxies and not being contaminated by other celestial objects.} GSMF, we construct the final galaxy sample by combining the KiDS internal morphological classification with external catalog cross-matching. A source is retained as a galaxy only if it simultaneously satisfies the following criteria: First, it must not be classified as a potential star in the KiDS DR4 catalog (\nolinkurl{SG2DPHOT}=0) \cite{Kuijken2019}. Second, after cross-matching with Gaia DR3 \citep{Delchambre2023} using a 1 arcsec radius, we examine it following a strict priority logic: we first rely on the discrete classification candidate flags (where \nolinkurl{in_qso_candidates} overrides \nolinkurl{in_galaxy_candidates}), and for the remaining unflagged sources, we assign the class with the highest probability among \nolinkurl{classprob_dsc_combmod_star}, \nolinkurl{classprob_dsc_combmod_galaxy}, and \nolinkurl{classprob_dsc_combmod_quasar}, which are the probabilities a source would be classified as a star/galaxy/quasar in Gaia DR3 Discrete Source classifier (DSC \cite{Bailer-Jones2021,GaiaCollaboration2023a}) system. An object is removed if determined to be a star or a QSO. Finally, we exclude any source that yields a match within 1 arcsec in the MILLIQUAS \citep{Flesch2023} catalog, as it is considered a QSO. The remaining sources constitute our final pure galaxy sample. In our work, following common practice in previous observational studies (e.g. \cite{Weaver2023,Davidzon2017}) we prioritize a conservative, contamination-minimized GSMF and exclude galaxies with QSO from our sample, a contribution which  appear small for the GSMF $< 0.1$ dex and relevant only for extremely high mass objects. 

    \begin{figure}[H]
        \centering
        \includegraphics[width=\columnwidth]{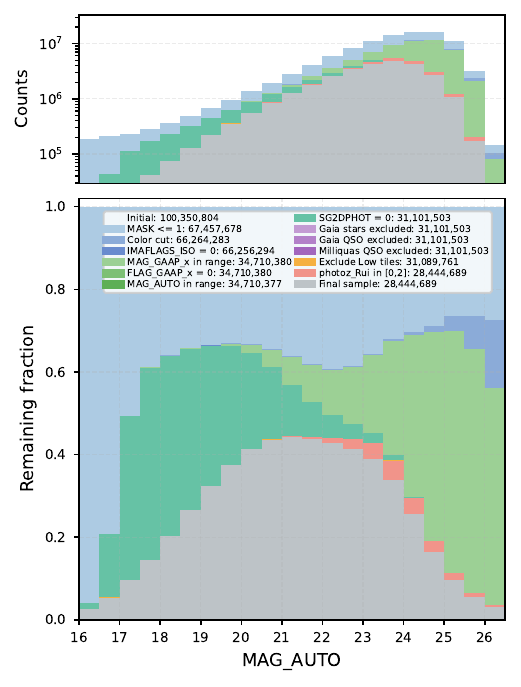}
        \caption{Sample reduction as a function of $r$-band \texttt{MAG\_AUTO}. The upper panel shows the number of sources in each magnitude bin on a logarithmic scale, with the overlapping coloured bars indicating how the distribution changes after successive screening steps. The lower panel shows the corresponding remaining fraction in each magnitude bin after each selection cut, illustrating which cuts dominate the sample reduction.}
        \label{fig:sample_reduction}
    \end{figure}

    Figure~\ref{fig:sample_reduction} summarizes how the source counts and remaining fraction evolve through our sample-selection pipeline as a function of \texttt{MAG\_AUTO}. We find that the \texttt{FLAG\_GAAP\_x} reduction has essentially no additional effect after the \texttt{MAG\_GAAP\_x} reduction, but we keep this step to preserve the consistency of the overall pipeline. We also note that the KiDS \texttt{SG2DPHOT} classification already removes most of the star/QSO contaminants that are later identified through Gaia and MILLIQUAS cross-matching. Nevertheless, we keep the Gaia/MILLIQUAS reduction both to maintain the same pipeline structure and to provide an external validation of our classification procedure. Among the selection steps, the MASK reduction produces the largest reduction across the full \texttt{MAG\_AUTO} range. The most influential magnitude-dependent reduction is caused by the 9-band selection, which mainly affects faint sources with \texttt{MAG\_AUTO}$>21$, while the KiDS galaxy classification primarily reduces bright sources with \texttt{MAG\_AUTO}$<23$.

    \subsection{Parameter estimation}
    \label{sec:parameter_estimation}
    
    \subsubsection{Photometric Redshift Estimation}
    \label{sec:redshift_estimation}

    The photometric redshift estimation is an important input in the SED fitting process, as it provides the era the detected light is coming from and this  information is needed to interpret the observed photometry in terms of physical properties of galaxies. 
    
    Following \citet{Xielh2023} we use the machine learning method GaZNet-z. GaZNet-z combines $r$-band imaging morphology with 9-band photometric data to estimate the redshifts of galaxies within our sample. This efficient method yields more accurate photometric redshift estimates than other traditional approaches \citep{Li2022} which also focus on photometry, albeit with limitations for some individual objects at high redshifts. In our work, we use GaZNet-z out to $z = 2$.

\begin{figure}[H]
        \centering
        \includegraphics[width=\columnwidth]{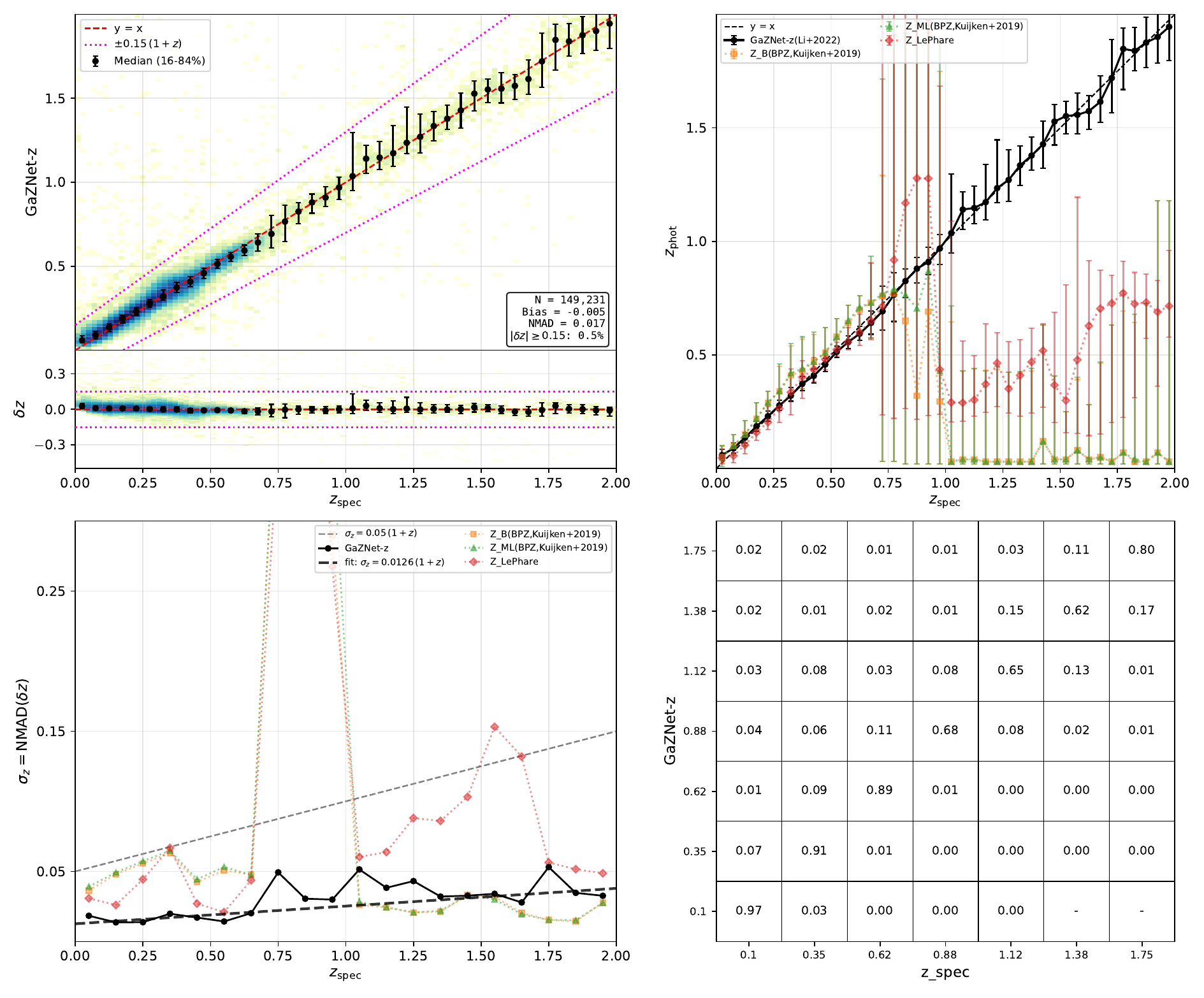}
        \caption{Photometric-redshift validation for GaZNet-z against spectroscopic redshifts. Top-left panel: two-dimensional density distribution of GaZNet-z versus $z_{\rm spec}$, with the red dashed line marking the one-to-one relation and the magenta dotted lines indicating the outlier threshold $|\delta z| = 0.15$. The black points show the median GaZNet-z values in $z_{\rm spec}$ bins with the 16--84\% range. The inset reports the sample size, bias, NMAD, and outlier fraction; the outlier fraction is only 0.5\%. Lower sub-panel in the top-left: normalized redshift residuals, $\delta z = (z_{\rm phot}-z_{\rm spec})/(1+z_{\rm spec})$, as a function of $z_{\rm spec}$. Top-right panel: row-normalized probability matrix showing the fraction of galaxies in each GaZNet-z bin that fall into each $z_{\rm spec}$ bin. Bottom-left panel: comparison of median photometric redshifts from GaZNet-z, Z\_B, Z\_ML, and Z\_LePhare as a function of $z_{\rm spec}$ \citep{Kuijken2019}. GaZNet-z shows better agreement with the one-to-one relation, especially at intermediate and high redshifts. Bottom-right panel: redshift dependence of $\sigma_z = {\rm NMAD}(\delta z)$ for GaZNet-z and the comparison methods. The gray dashed line shows the typical relation $\sigma_z = 0.05(1+z)$ from \citet{Newman2022}, while the black dashed line shows the best-fit trend for GaZNet-z.}
        \label{fig:z-comparison}
    \end{figure}

    Figure~\ref{fig:z-comparison} illustrates the performance of GaZNet-z against spectroscopic redshifts using four diagnostic sub-panels. The spectroscopic redshifts are from GAMA\footnote{https://www.gama-survey.org/dr4/schema/table.php?id=684} with excluding stars (by using $SC\geq2$). In the top-left panel, the main panel compares GaZNet-z with $z_{\rm spec}$ through a two-dimensional density map, while the lower residual panel shows $\delta z = (z_{\rm phot}-z_{\rm spec})/(1+z_{\rm spec})$ as a function of $z_{\rm spec}$. The reported bias is defined as $\mathrm{median}(z_{\rm spec}-z_{\rm phot})$, NMAD is computed as $1.4826\,\mathrm{median}(|\delta z-\mathrm{median}(\delta z)|)$, and outliers are defined by $|\delta z|\geq 0.15$. The resulting outlier fraction is only 0.5\% for the total sample, lower than the values reported by \citet{Tomczak2014} (3\%), \citet{Ilbert2013} (2.1\%) and \citet{Weibel2024} (5\%), and the median GaZNet-z trend remains close to the one-to-one relation, especially where spectroscopic data are abundant at $z<1$.

\end{multicols}

\begin{multicols}{2}
    
    In the top-right panel we compare the median photometric-redshift trends for GaZNet-z, Z\_B, Z\_ML, and Z\_LePhare, showing that GaZNet-z follows well the one-to-one relation and typically out-performs  traditional photometric-redshift estimates. We want to emphasize that our test employs a 9-band dataset and essentially it is an assessment of how a redshift determination technique performs upon having moderate wavelength coverage. LePhare has been shown to have  relatively also a good redshift determination, however upon data-sets with considerable wavelength (many bands) coverage.

   In the bottom-left panel we summarize the redshift dependence of $\sigma_z=\mathrm{NMAD}(\delta z)$. We fit the GaZNet-z trend as $\sigma_z = 0.0126(1+z)$ and compare it with the commonly adopted SED-photometric-redshift performance from \citet{Newman2022}, $\sigma_z = 0.05(1+z)$. The lower normalization of the GaZNet-z relation demonstrates its substantially smaller redshift scatter over the range considered here. We have to note though that our scatter is calculated using as a reference  the spectroscopic redshift and is not an error derived from GaZNet-z \footnote{GaZNet-z has the limitation that it cannot provide errors on the derived z as a ML method that adopts a convolution neural network.}. 

In the bottom-left panel we provide  the row-normalized probability matrix used to evaluate the redshift-bin mixing due to the uncertain redshift determination from GaZNet-z. For each GaZNet-z bin, the entries show the fraction of galaxies whose true spectroscopic redshifts fall into each $z_{\rm spec}$ bin. This panel is an interesting  accuracy summary as it shows some interesting limitations at the intermediate redshifts considered in our work. At the z = 0.1 bin 97\% of our galaxies from GaZNet-z originate from galaxies that belong at  the same spectroscopic redshift bin. However, going to the redshift bin of 1.38 we see that the fraction decreases to 62\%. This shows that GaZNet-z, even if it is a considerable improvement with respect traditional methods (that would give a very concerning matrix) there are still limitations that have to be acknowledged. The matrix indicates which spectroscopic-redshift bins contribute to a given GaZNet-z estimate, and can subsequently be used to characterize and correct the GaZNet-z scatter/migration between redshift bins something that as we will see in subsection \ref{subsec:ScatterDueToUncertaity} can impact the normalization of the GSMF by $\sim $ 0.1 dex.

    We make use of $0.001<z<2$ sources, with this redshift cutting and combining sample selection in Section ~\ref{sec:sample_construction}, we finally get $28M$ galaxies that will be used to construct GSMFs at different redshifts. The detailed source reduction is showed in Fig~\ref{fig:sample_reduction}.

\begin{figure}[H]
        \centering
        \includegraphics[width=\columnwidth]{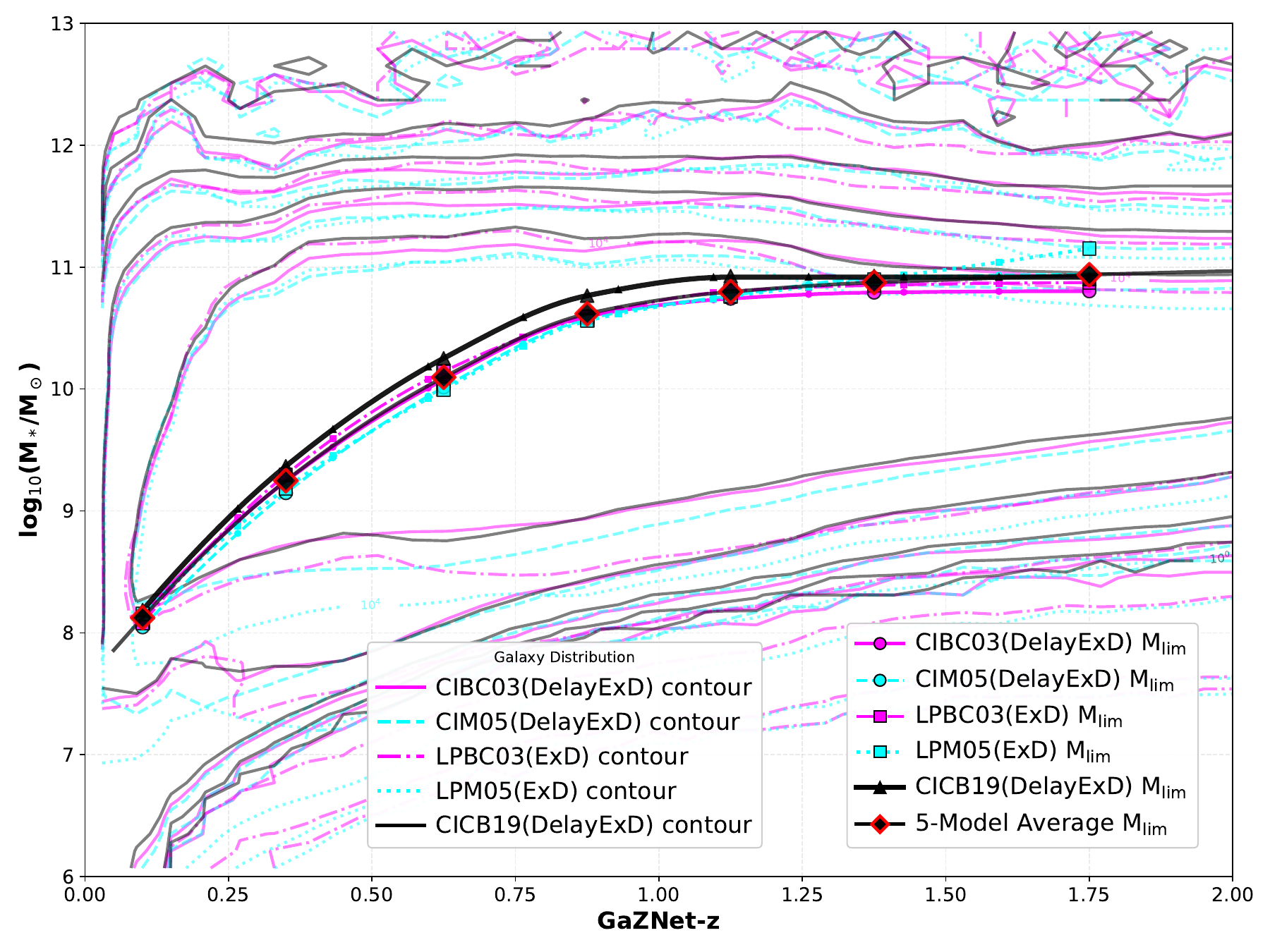}
        \caption{Stellar mass completeness as a function of redshift for the KiDS galaxy sample in the $\log M_*/M_{\odot}$--$z$ plane. Contours show the two-dimensional galaxy distribution for the SED-fitting models, with line styles and markers distinguishing CIBC03, CIM05, LPBC03, LPM05, and CICB19 (solid circles, dashed circles, dash-dot squares, dotted squares, and solid triangles, respectively). Colored lines with symbols show the individual $M_{\rm lim}(z)$ estimates from each model. The black line with red-edged diamonds shows the 5-model average completeness limit adopted for the GSMF analysis. Contour labels indicate $\log_{10}$ of the number of galaxies.}
        \label{fig:smf_completeness}
    \end{figure}

\begin{table*}[t]
    \centering
    \caption{Comparison of SED fitting model parameters used in this study.}
    \label{tab:model_parameters}
    
    \small
    \begin{tabular}{@{} l *{5}{C{0.15\textwidth}} @{}}
        \toprule
        \textit{Parameter}  
        & \textit{CICB19} 
        & \textit{CIBC03} 
        & \textit{CIM05} 
        & \textit{LPBC03} 
        & \textit{LPM05} \\
        \midrule
        
        Code                
        & \multicolumn{3}{C{0.54\textwidth}}{CIGALE} 
        & \multicolumn{2}{C{0.36\textwidth}}{LePhare} \\
        
        SFH                 
        & \multicolumn{3}{C{0.54\textwidth}}{Delayed Exp. Declining} 
        & \multicolumn{2}{C{0.36\textwidth}}{Exp. Declining} \\
        
        Templates           
        & CB19 
        & BC03 
        & M05 
        & BC03 
        & M05 \\
        
        IMF                 
        & Chabrier 
        & Chabrier 
        & Kroupa 
        & Chabrier 
        & Kroupa \\

        \makecell{$Z$ ($10^{-2}$)}     
        & 0.01, 0.02, 0.05, 0.1, 0.2, 0.4, 0.6, 0.8, 1.4, 1.7, 1, 2, 3, 4, 6 
        & 0.01, 0.04, 0.4, 0.8, 2, 5 
        & 0.1, 1, 2, 4 
        & 0.4, 0.8, 2 
        & 0.4, 1, 2 \\
        
        $\tau$              
        & \multicolumn{3}{C{0.54\textwidth}}{0.25, 0.5, 0.75, 1, 1.5, 2, 2.5, 3, 4, 5} 
        & \multicolumn{2}{C{0.36\textwidth}}{0.1, 0.3, 1, 2, 3, 5, 10, 15, 30} \\

        age/Gyr             
        & \multicolumn{3}{C{0.54\textwidth}}{0.5, 1, 1.5, 2, 2.5, 3, 4, 5, 6, 8, 10, 13} 
        & \multicolumn{2}{C{0.36\textwidth}}{0.5--13.5} \\
        
        Dust Law            
        & \multicolumn{5}{C{0.9\textwidth}}{Calzetti et al. (2000)} \\ 
        
        $E(B-V)$            
        & \multicolumn{3}{C{0.54\textwidth}}{0.01, 0.02, 0.04, 0.06, 0.1, 0.2, 0.4, 0.6, 0.8} 
        & \multicolumn{2}{C{0.36\textwidth}}{0, 0.1, 0.2, 0.3} \\
        
        \bottomrule
    \end{tabular}
\end{table*}

    \subsubsection{Stellar Mass Estimation}
    \label{sec:stellar_mass_estimation}

We derive stellar masses using the SEDs of galaxies, which are fitted to a library of templates that represent different stellar populations to get some key parameters, including stellar mass and star formation rates.
    \textit{SED Fitting configuration:}
    Since galaxy properties derived from SED fitting are sensitive to model assumptions we adopt five configurations by combining two fitting codes (\texttt{CIGALE \cite{Boquien2019}} and \texttt{LePhare \cite{Arnouts1999,Ilbert2006}}; therefore star formation histories, delayed exponential declining SFH and exponential declining SFH, respectively) with three SPS template libraries (CB19 \citep{CB19}, BC03 \cite{Bruzual2003} and M05 \cite{Maraston2005}). In these runs, BC03+CB19-based configurations adopt a Chabrier IMF \cite{Chabrier2003}, while M05-based configurations adopt a Kroupa IMF \cite{Kroupa2001}; and all configurations use the Calzetti dust attenuation law \citep{Calzetti2000}. We consistently model and compare the impact of these setup choices, together with the adopted metallicity, age, and reddening grids, to quantify systematics associated with code/template choices within a uniform stellar-mass estimation framework. Detailed parameter values for each configuration are listed in Table~\ref{tab:model_parameters}.

     \textit{Stellar Mass Correction:} When estimating the stellar mass from SED fitting, we account for the aperture effects following \citet{Xielh2023}, \citet{Bilicki2021}, \citet{Taylor:2011dk}, \citet{Sureshkumar2024a} and  apply a correction factor based on the difference between the total magnitude and the aperture magnitude. The correction factor is given by:
    \begin{equation}
        \label{eq:aper_corr}
        \begin{aligned}
        \log M_{*, \mathrm{corr}}/M_{\odot} &= \log M_{*, \mathrm{SED}}/M_{\odot} \\
        &\quad + 0.4 \times (MAG\_AUTO - MAG\_GAAP\_r)
        \end{aligned}
    \end{equation}
        where $\log M_{*, \mathrm{corr}}/M_{\odot}$ is the corrected stellar mass, $\log M_{*, \mathrm{SED}}/M_{\odot}$ is the stellar mass estimated from the SED fitting, $MAG\_AUTO$ and $MAG\_GAAP\_r$ are respectively the total and aperture magnitude of the $r$-band of the galaxy (see Section~\ref{sec:base_to_9band}). While the application of an r-band fluxscale correction is standard practice in the literature, we acknowledge that this approach may not fully account for redshift-dependent aperture effects \cite{Paulino-Afonso2022,Zibetti2026,Zhu2023} 

    The resulting catalog contains 28 million galaxies with \textit{r}-band magnitudes between 15 and 26, providing a robust dataset for constructing the GSMF across a wide range of stellar masses and redshifts. The catalog will be made publicly available. A summary of the catalog columns is provided in the Supplementary Materials, where the catalog is referred to as KiDSDR4-C0v1.1.
    \subsection{Stellar-mass-complete sample}
    \label{sec:completeness}
    The incompleteness of our  stellar mass  arises because galaxies exhibit an intrinsic scatter in their mass-to-light ($M_*/L$) ratios, making systems with high $M_*/L$ more difficult to detect at a fixed flux limit (a kind of Malmquist bias \cite{Butkevich2005}). We apply the empirical method described in \citep{Marchesini2009,Pozzetti2010}. This completeness threshold is subsequently utilized in Section~\ref{sec:vmax} to appropriately correct the observations at the low-mass end.

    To estimate the completeness, we used the following steps:
    \begin{itemize}
        \item We calculated $\log(\mathrm{sSFR})/\mathrm{yr}^{-1}$ for each galaxy in the sample using the SED fitting results. 
        We adopted a passive-galaxy threshold of $\mathrm{sSFR}/\mathrm{Gyr}^{-1} < 10^{-1.8+0.3\times z}$, following \citet{Dave2019} and use this criterion consistently in our completeness estimate.
        \item We calculated the stellar mass limit for each galaxy based on the following formula: $\log M_{\mathrm{lim}}/M_{\odot} = \log M_{*, \mathrm{corr}}/M_{\odot}+0.4\times(MAG\_GAAP\_r-MAG\_GAAP\_r_{\mathrm{lim}})$ where $\log M_{*, \mathrm{corr}}$ is the aperture-corrected stellar mass estimated from the SED fitting(see equation~\ref{eq:aper_corr}),  and $MAG\_GAAP\_r_{\mathrm{lim}}$ is the $r$-band magnitude limit of the survey, we took the most loose one, equaling to 25 mag \cite{Kuijken2019}. This relation rescales each galaxy to the survey flux limit under fixed $M_*/L$, i.e., it estimates the minimum stellar mass that the same object would have if observed exactly at the magnitude limit.
        \item We selected the faintest 20\% sources and then determined the stellar mass limit below which lay 95\%  sources of these faint objects \cite{Weigel2016}. This percentile-based envelope defines $M_{\mathrm{lim}}(z)$ for each SED setup and is then used as the redshift-dependent completeness cut in all subsequent GSMF measurements (vertical completeness cuts and retained mass bins).
    \end{itemize}
    In Figure~\ref{fig:smf_completeness} we present the stellar mass completeness as a function of redshift for  the KiDS galaxy sample. The following results and analysis  all consider the mass limitations.

        \section{GSMF Estimation Method}
    \label{sec:GSMF-methods}
    \subsection{Selection Effects: Inverse Probability Selection Weighting}
    \label{sec:IPSW}

    To account for selection effects arising from the requirement of complete 9-band photometry for SED fitting, we applied Inverse Probability Selection Weighting \citep[IPSW;][]{Horvitz1952}. This method corrects for selection effects by weighting each galaxy in the selected sample ($S$, 9-band sample in this work) by the inverse of its inclusion probability from the base sample ($U$, base sample in this work).

    We modeled the selection probability $\pi_i = P(i \in S | X_i)$ using a logistic regression model. The model utilized four primary auxiliary features $X_i$ available in the base sample \cite{Kuijken2019}:
    \begin{itemize}
        \item \texttt{MAG\_AUTO}: see Section~\ref{sec:base_to_9band}, representing the total magnitude of the galaxy;
        \item $Z\_B$: The nine-band BPZ photometric redshift estimate,as a proxy for redshift information;
        \item \texttt{COLOUR\_GAAP\_g\_r}: Optical $g-r$ color;
        \item \texttt{COLOUR\_GAAP\_Y\_J}: Near-infrared $Y-J$ color.
    \end{itemize}
    
    As the selection of the 9-band sample is based on the availability of photometry across all bands, these features are expected to capture the key dependencies of the selection process, including photometry, redshift, and color information that influence the likelihood of a galaxy being included in the 9-band sample. To handle missing data, we included binary indicators for missing values and applied median imputation. All features were standardized before training.

    Following the principle of propensity score calibration \citep{Rosenbaum1983}, we implemented a logit offset constraint to ensure the sum-to-universe condition:

    \begin{equation}
        \sum_{i \in S} w_{IPSW,i} = N_U
    \end{equation}
    where $\text{logit}_i \equiv \ln\!\left(\frac{\pi_i}{1-\pi_i}\right)$ and the weights are defined as $w_{IPSW,i} = 1 + \exp(-(\text{logit}_i + k)) = 1/\pi'_i$, with $\pi'_i$ the shifted propensity after adding an intercept offset $k$ in logit space. The optimal offset $k$ was determined numerically so that $\sum_{i\in S} w_{IPSW,i}$ matches $N_U$. In other words, $k$ is a global calibration parameter that preserves the relative ranking of selection probabilities while correcting the overall normalization of the weights to the parent catalog. We note that these weights estimate the selection effect relative to the base catalog's composition, which also includes stars/quasars.

    \subsection{1/Vmax method}
    \label{sec:vmax}

    We used the 1/Vmax method \citep{Schmidt1968} to estimate the GSMF of galaxies in the KiDS DR4. The 1/Vmax method is a non-parametric approach that allows us to estimate the GSMF without assuming a specific functional form for the distribution of stellar masses. 

    The 1/Vmax method works by calculating the maximum comoving volume available for each galaxy in the survey, which is determined by the redshift range and the survey area.  For each galaxy, we calculate the maximum redshift $z_{max}$ at which it could still be detected given the survey's magnitude limits and stellar mass completeness.

The GSMF is then estimated by summing the contributions of each galaxy to the mass function, weighted by the inverse of the maximum volume in which they could be detected.

        To further refine the GSMF estimate, we incorporate the IPSW weights $w_{IPSW}$ derived in Section~\ref{sec:IPSW}. The updated formula for the GSMF is:
    \begin{equation}
        \Phi(M) = \frac{1}{\Delta M} \sum_{i} \frac{w_{IPSW,i}}{V_{max,i}},
    \end{equation}
    where $w_{IPSW,i}$ is the inverse probability selection weight for galaxy $i$, and $V_{max,i}$ is the maximum comoving volume.

    The maximum comoving volume for each galaxy is calculated as follows \citep{Hogg2000,Weigel2016}:
    \begin{equation}
        V_{max,i} = \frac{4\pi}{3} \times \frac{\Omega_{survey}}{4\pi} \times [d_c(z_{max,i})^3 - d_c(z_{min})^3]
    \end{equation}

    where $\Omega_{survey}$ is the effective survey area, $d_c(z)$ is the comoving distance to redshift $z$, $z_{max,i}$ is the maximum redshift at which galaxy $i$ could be detected (limited by either the redshift bin boundary or the stellar mass completeness limit), and $z_{min}$ is the minimum redshift of the redshift bin.
    
    For error estimation, we adopt a modified approach based on effective number and weight statistics \citep{Gehrels1986}. For each mass bin, we calculate the effective weight $W_{e}$ and effective number $N_{e}$ as:
    \begin{equation}
        W_{e} = \frac{\sum_i w_i^2}{\sum_i w_i},N_{e} = \frac{\sum_i w_i}{W_{e}}
    \end{equation}

    where $w_i = w_{IPSW,i}/V_{max,i}$ is the total weight for galaxy $i$. We then use a modified Gehrels approach with 84\% confidence level (equation(7) and (11) in \citet{Gehrels1986}, and like \citet{Weigel2016} we took $S = 1$) to estimate the asymmetric errors:
    \begin{equation}
        \begin{split}
            \sigma_{u} = \frac{W_{\text{eff}}}{\Delta M} \left( N_{\text{eff}} + \sqrt{N_{\text{eff}} + \frac{3}{4}} + 1 \right) - \Phi(M)
        \end{split}
    \end{equation}

    \begin{equation}
        \begin{split}
            \sigma_{l} = \Phi(M) - \frac{W_{\text{eff}}}{\Delta M} \left( N_{\text{eff}} - \sqrt{N_{\text{eff}} - \frac{1}{4}}  \right)
        \end{split}
    \end{equation}

\subsection{Scatter Due to GaZNet-z uncertainty.}
\label{subsec:ScatterDueToUncertaity}

The uncertainty in photometric redshifts can propagate into the accessible volume, $V_{\max}$, and therefore into the normalization of the GSMF. For GaZNet-z, however, we do not have a direct redshift uncertainty estimate for each individual source, so a source-by-source error propagation into $V_{\max}$ is not possible. Instead, we use the bottom-right panel of Figure~\ref{fig:z-comparison}, which gives the row-normalized probability matrix $P(z_{\rm spec}\,\mathrm{bin}\,j\,|\,z_{\rm phot}\,\mathrm{bin}\,i)$. This matrix quantifies, for a fixed GaZNet-z bin, the fraction of galaxies that statistically originate from each spectroscopic-redshift bin, and therefore provides an estimate of redshift-bin migration.

We use this probability matrix to estimate a bin-level normalization correction to the GSMF. For each GaZNet-z bin $i$, we first count the nominal number of galaxies assigned to that bin, $N_i^{\rm phot}$, and then redistribute these counts across the true-redshift bins using the probability matrix, such that $N_j^{\rm true}=\sum_i P_{ij}N_i^{\rm phot}$. The correction factor for each redshift bin is then defined as the ratio between this expected true number and the nominal GaZNet-z number in the same bin, using the same weighting convention in the numerator and denominator. We apply this estimate after the same galaxy-selection and mass-completeness cuts used for the GSMF measurement, and also consider the IPSW-weighted counts used in the final analysis. This procedure does not correct individual galaxy redshifts, but it captures the net effect of GaZNet-z scatter on the overall GSMF normalization in each redshift bin. The resulting normalization shifts are modest, typically of $\sim 0.1$ dex across the adopted redshift bins. More details on this correction can be found in the Supplementary Materials.

\subsection{The redshift uncertainties for high masses.}
\label{subsec:redshiftuncertantiesMass}

Before comparing our reference configuration with other studies in subsection \ref{sec:smf_lits}, it is important to remind that redshift determination can affect the GSMF, especially at the high mass end. In section \ref{sec:redshift_estimation} we demonstrated that GaZNet-z typically outperforms previous SED based photometric methodologies. However, sometimes a massive galaxy can be the result of a flawed redshift determination. For example a close object with high luminosity can be placed at high redshift and this would result in inferring a high mass for the galaxy. 

\begin{figure}[H]
        \centering
        \includegraphics[width=\columnwidth]{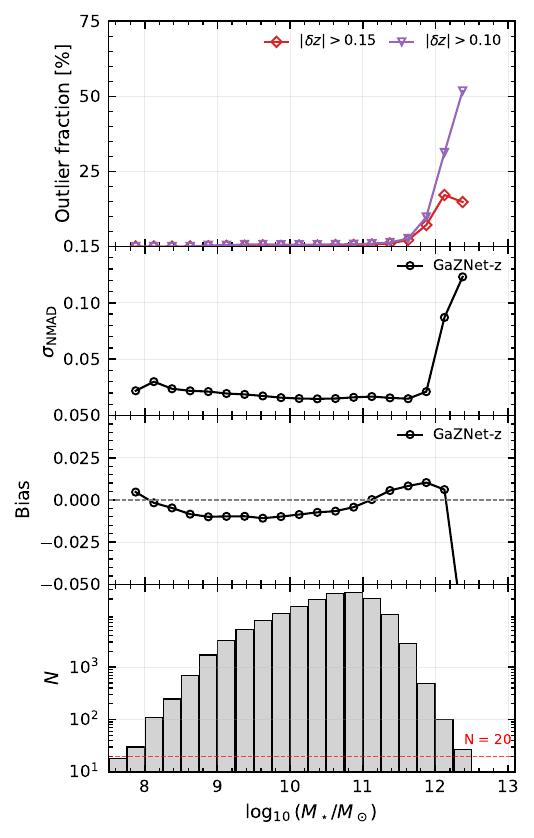}
        \caption{Dependence of GaZNet-z redshift-performance metrics on stellar mass. From top to bottom, the panels show the outlier fraction for $|\delta z|>0.15$ and $|\delta z|>0.10$, the NMAD scatter $\sigma_{\rm NMAD}$, the redshift bias, and the number of spectroscopic sources in each stellar-mass bin. Only bins with at least 20 sources are used for the redshift-performance statistics; the bottom panel shows this threshold as a dashed red line.}
        \label{fig:z-comparisonMassive}
    \end{figure}

Thus, we estimate what fraction of massive objects can be the result of misguided redshift estimation by studying the dependence of the outlier fraction with mass in Figure~\ref{fig:z-comparisonMassive}, using the spectroscipic redshifts we used in Figure~\ref{fig:z-comparison}

    In the top panel of Fig. \ref{fig:z-comparisonMassive} we study the relation of the outlier fraction with mass. If we adopt a catastrophic outlier fraction threshold $|\delta z| = 0.15$ and focus specifically only on the most massive bins the outlier fraction is $\sim$ 17$\%$. If we adopt an even stricter outlier fraction threshold criterion of 10$\%$ the outlier fraction with bad redshift estimation would reach values of 50$\%$ for the highest mass bin.  This limitation in deriving redshifts for some massive objects should be taken into account and we remove a substantial fraction of objects based on the $\delta z>$ 0.10 criterion from the most massive bins of our GSMF. In the bottom panel we show the histogram with respect stellar mass, we noted that we don't have $\log M_*/M_{\odot}>12.5$, which means we can't constrain our GaZNet-z up to that stellar range while the number of sources in KiDS DR4 catalog with that mass are few, we decide to exclude these objects.
    
Thus, in our work we decrease our  GSMF of the most massive bins with $\log M_*/M_{\odot}=11.875, 12.125, 12.375$   by $\sim$ $0.045 , 0.163 , 0.317~dex$.     For lower masses than $< 10^{11.75}$ $M_\odot$ the outlier fraction is small so we do not consider a correction. More details on this correction can be found in the Supplementary Materials.

    \section{GSMF results and discussion}
    \label{sec:GSMF-results-discussion}
\begin{figure*}[t]
    \centering
    \includegraphics[width=0.95\textwidth]{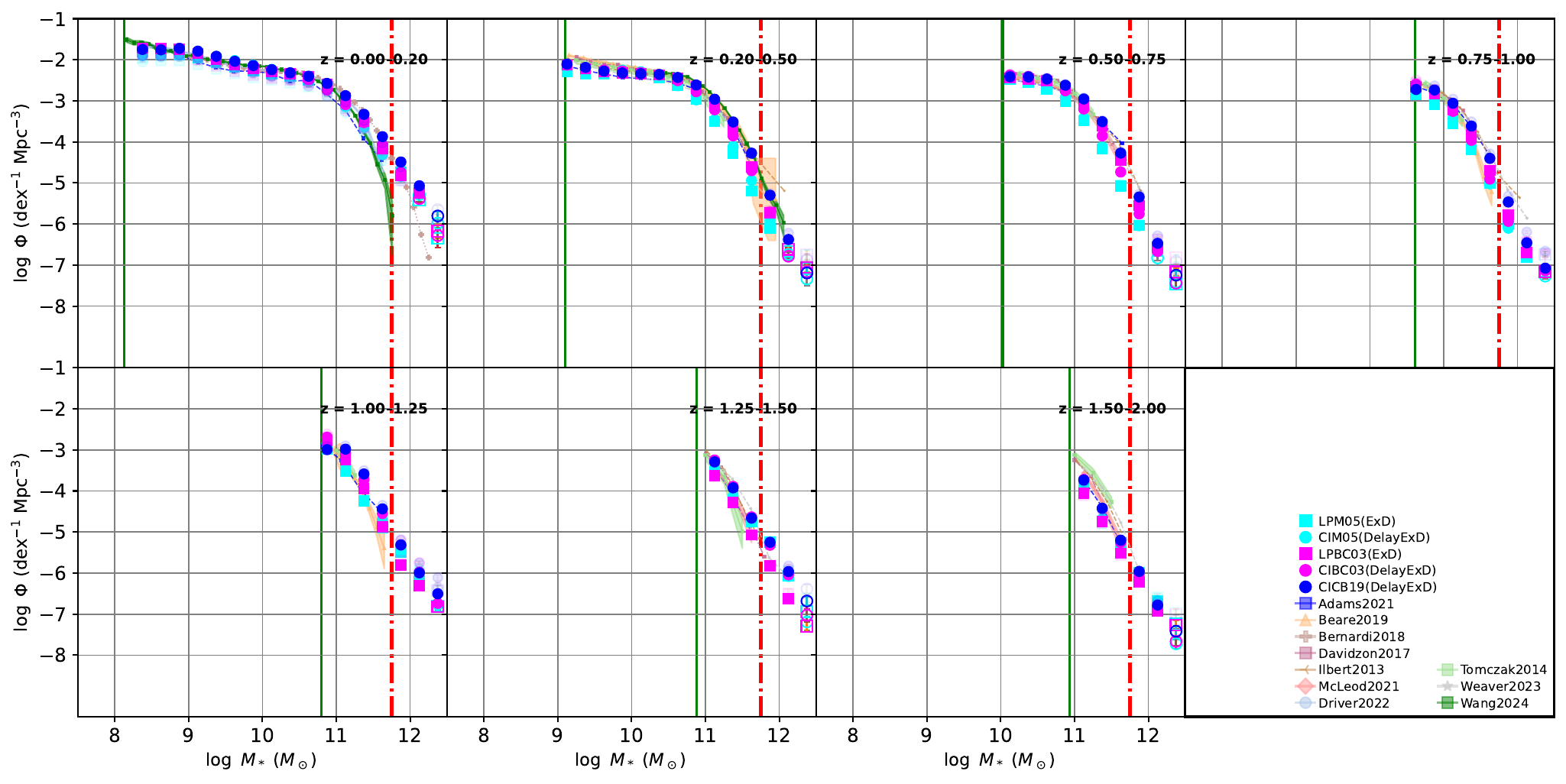}
    \caption{GSMF comparison across redshift bins, showing KiDS DR4 results for five SED-fitting configurations and selected literature measurements. For KiDS DR4, symbols: CICB19 (blue circles), CIBC03 (magenta circles), CIM05 (cyan circles), LPBC03 (magenta squares), and LPM05 (cyan squares); the thick black curve/symbols show the four-model mean. Literature datasets are shown with distinct colors/markers(Adams2021: blue squares; Beare2019: orange triangles; Bernardi2018: brown pentagons; Davidzon2017: purple squares; Ilbert2013: brown triangles; McLeod2021: pink diamonds; Driver2022: light-blue circles; Tomczak2014: green squares; Weaver2023: gray stars;  Wang2024: green squares).}
    \label{fig:smf_results}
\end{figure*}

   \begin{figure*}[t]
    \centering
    \includegraphics[width=\textwidth]{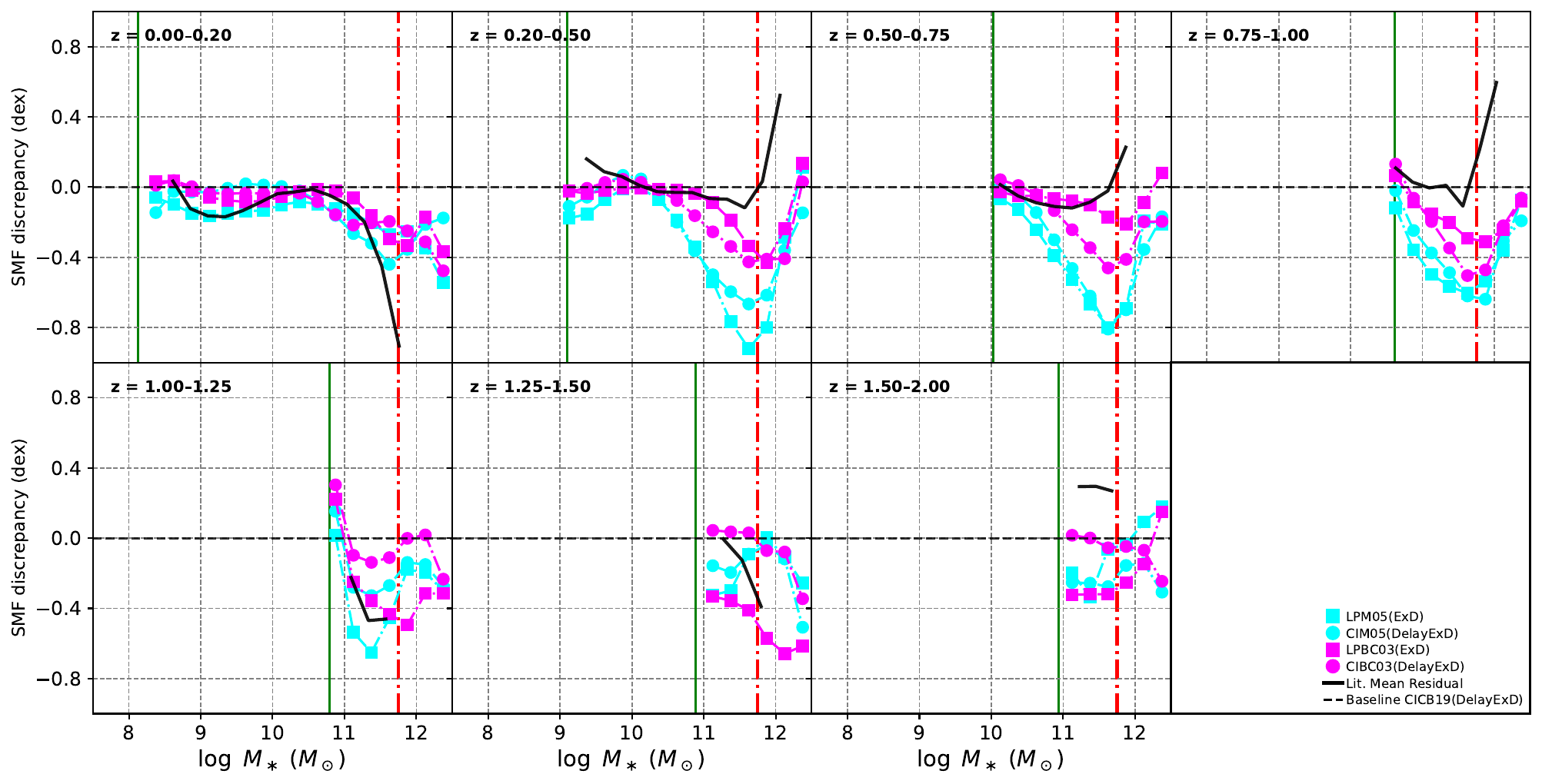}
    \caption{Residuals of the GSMF with respect to the four-model mean in each redshift bin. KiDS DR4 configurations and literature datasets use the same symbol/color convention as Figure~\ref{fig:smf_results}. The black solid curve is the mean of residual of literature compiled in each redshift bin, compiled in table ~\ref{tab:literature} and Figure~\ref{fig:smf_literature_residuals}.}
    \label{fig:smf_residuals}
\end{figure*}

    \subsection{GSMF results of different SED prescriptions}
    \label{sec:smf_codes_residuals}

    In SED fitting process, a key source of systematics is the choice of SPS library. This is due to the fact that different libraries include different ingredients for stellar populations. For example, the M05 templates are based on the fuel-consumption theorem and rely on Cassisi + Geneva evolutionary tracks (without overshooting), explicitly modeling phases such as the horizontal branch (HB) and thermally-pulsing asymptotic giant branch (TP-AGB). In contrast, BC03 is based on isochrone synthesis, commonly using Padova 1994 tracks with overshooting, and provides a comparatively simpler treatment of the HB/TP-AGB phases. Compared with BC03, CB19 includes updated treatments for massive stars, including Wolf-Rayet stars, and extends the stellar-population modelling to the corresponding metallicity regimes where such massive-star phases are important. These differences among SPS can change the predicted mass-to-light ratios and therefore shift stellar masses in a redshift- and age-dependent way.

    Following \citet{Xielh2023} we use the 5 different configuration to test how our results change upon different assumptions. In our configurations, BC03+CB19 runs adopt a Chabrier IMF, while M05 runs adopt a Kroupa IMF (Table~\ref{tab:model_parameters}). We therefore apply a $-0.05$ dex correction when comparing M05-based stellar masses to the Chabrier scale.

        The main difference between CIGALE and LePhare configurations in our work is the assumed SFH functional form and its parameter sampling. Larger $\tau$ values correspond to a more slowly declining SFH (i.e., a more extended star-formation timescale and relatively higher late-time SFR). CIGALE adopts a delayed exponentially-declining SFH, while LePhare uses a simple exponentially-declining SFH and explores a wider $\tau$ grid.

     Beyond SFH, CIGALE also samples a wider and higher-resolution grid in dust attenuation $E(B-V)$ and metallicity than LePhare in our setup (Table~\ref{tab:model_parameters}); in this context, the smaller residuals seen for CIGALE can plausibly reflect its broader parameter coverage (together with its energy-balance constraint), which reduces template-mismatch systematics.

    In Figure~\ref{fig:smf_results} we present our GSMF results obtained from the 5 SED-fitting configurations. At first glance the different configurations do not show apparent differences. However, in Figure~\ref{fig:smf_residuals} we present the residuals of the different setups with respect to the results of our reference configuration that adopts a CB19 SPS for clarity. The comparison illustrates that the inferred GSMF is relatively stable against the choice of code/SFH at the level of the statistical uncertainties, but systematic trends become apparent when splitting the configurations by stellar population synthesis model (i.e. CB19 vs BC03 vs M05).

    The residuals of our different configruations are smallest in the lowest-redshift bin ($z=0$--0.2), as expected given that SPS libraries and attenuation prescriptions are largely both  constrained at low redshift (Table~\ref{tab:model_parameters}). We note though that typically adopting a CB19 SPS,  results in larger values for the high mass end, especially for objects with masses of $> 10^{11.5}$ $M_\odot$. 
    
    At $z=0.2$--1.0, the BC03-based and CB19-based configurations lie systematically above the M05-based configurations (for $\log(M_*/M_\odot)\gtrsim 10$). The mass-dependent trend shows a ``peak--valley'' feature around $\log(M_*/M_\odot)\sim 11.5$: Both BC03 and M05 residuals typically decline, reach a minimum and then rise again with increasing stellar mass to atain values close to our reference CB19 model. We note that the the maximum offset introduced by choice of SPS at this redshift interval can reach even 0.9 dex.  

    A plausible interpretation is that the more detailed treatment of late evolutionary phases in M05 (i.e. those of HB and TP-AGB; fuel-consumption theorem; Cassisi+Geneva tracks without overshooting) can change near-IR mass-to-light ratios relative to BC03 (Padova 1994 with overshooting and a weaker HB/TP-AGB description), leading to systematically lower best-fit stellar masses for the same photometry and thus lower inferred number densities at higher fixed $M_*$. Consistent with this, across $0<z<2$ the BC03 and CB19 are both almost always $\ge$ those of M05.

    Differences between the two fitting codes combined with the SFH choices and parameter grids are also evident when we focus the comparison  solely on the BC03 (i.e. CICB03 versus LPBC03).  We note that beyond SFH, CIGALE also samples a wider and higher-resolution grid in dust attenuation $E(B-V)$ and metallicity than LePhare in our setup (Table~\ref{tab:model_parameters});  At $z<1$ in the mass-complete regime, Lephare and CIGALE behavior is similar when considering the same SPS. However, At $z=1.0$--1.5, the separation bettween them becomes more apparent and can reach even 0.4 dex.

    At $z=1.5$--2.0, the four configurations show more complex mass-dependent behavior with respect our reference. However, the residual decreases again to negative for very massive objects like at other redshifts. CIM05 and LPBC03 show opposite ``up--down'' versus ``down--up'' trends, and a clean separation of SPS- versus code-driven systematics is more difficult to infer. Nevertheless, overall the comparison suggests that both SPS effects and code/SFH effects remain relevant at the highest redshifts probed here.
\begin{table*}[t]
    \centering
    \caption{Key observational characteristics of literature studies used for GSMF comparison.}
    \label{tab:literature}
    \scriptsize
    \setlength{\tabcolsep}{4pt}
    \begin{tabular}{p{3cm}p{3.5cm}cp{2.8cm}ccc}
        \hline
        \hline
        Reference            & Survey                                                    & Area      & Primary Bands                                & Depth Limit           & Redshift       & Mass Method             \\
                             &                                                           & (deg$^2$) &                                              & (mag)                 & Type           &                         \\
        \hline
        \citet{Adams2021}    & COSMOS/XMM-LSS+ \newline SpizerIRAC                       & 5.23      & $u^*/g/r/i/z/y/Y/J/H/K_s$/ \newline IRAC     & $u^*=27.2$            & Photometric    & SED-LP                  \\
        \citet{Beare2019}    & Boötes+NDWFS/ \newline NEWFIRM/Miyazaki12/ \newline SDWFS & 8.26      & $B/J/H/K_s/z$/ \newline IRAC                 & $I=24$                & Photometric    & $M/L_K$                 \\
        \citet{Bernardi2018} & SDSS DR7                                                  & 4700      & $u/g/r/i/z$                                  & $r<17.77$             & Spectroscopic  & M/L                     \\
        \citet{Davidzon2017} & COSMOS2015                                                & 1.7       & GALEX+Optical+NIR+ \newline IRAC$^a$         & --                    & Photometric    & SED-LP                  \\
        \citet{Ilbert2013}   & UltraVISTA                                                & 1.52      & Optical+Medium/Narrow+ \newline NIR+IRAC$^b$ & $K_s<24$              & Photometric    & SED-LP                  \\
        \citet{McLeod2021}   & UKIDSS UDS+COSMOS+ \newline CFHTLS-D1                     & 2.57      & CFHT+VISTA+SSC+ \newline IRAC$^c$            & $B\sim27.4$           & Photometric    & SED-LP                  \\
        \citet{Driver2022}   & GAMA                                                      & 230       & UV-to-FIR$^d$                                & $r_{KiDS}=19.65$      & Spectroscopic  & SED-LP                  \\
        \citet{Tomczak2014}  & ZFOURGE/CANDELS                                           & 0.088     & HST+Ground-based$^e$                         & $J_{125}<26.5$        & Photometric    & SED-FAST                \\
        \citet{Weaver2023}   & COSMOS2020                                                & 1.27      & GALEX+Optical+NIR+ \newline IRAC$^f$         & $F814W\sim28$         & Photometric    & SED-LP                  \\
        \citet{Wang2024}     & DESI SV3-BGS                                              & 133       & g/r/z/W1/W2                                  & $r\leq19.5$+$z\leq19$ & Photo-z+Spec-z & SED-CIGALE/k-correction \\
        \hline
    \end{tabular}
    \\[0.2em]
    \footnotesize
    \textit{Notes:} $^a$GALEX FUV/NUV + MegaCam $u/u^*$ + HST F814W + HSC $grizy$ + Subaru bands + UltraVISTA $YJHK_s$ + IRAC. $^b$CFHT $u^*/B_J/V_J/r^+/i^+/z^+$ + Subaru bands + UltraVISTA $YJHK_s$ + IRAC. $^c$CFHT $u^*griz$ + VISTA $YJHKs$ + SSC $BVRiz'$ + WFCam $JHK$ + IRAC. $^d$GALEX UV + KiDS optical + VISTA VIKING NIR + WISE + Herschel. $^e$HST WFC3/ACS filters + Ground-based $YJHKs$ + narrow bands. $^f$Similar to COSMOS2015 with improved photometry. SED-LP: LePhare fitting; SED-FAST: FAST fitting; Color-M/L: Color-based mass-to-light ratio; M/L: Mass-to-light ratio. All studies assume $H_0=70$ km s$^{-1}$ Mpc$^{-1}$, $\Omega_m=0.3$, $\Omega_\Lambda=0.7$. All studies have used BC03 templates except \citet{Beare2019} and \citet{Barber2018}. All studies have used exponential declining SFH, except \citet{Weaver2023} and \citet{Driver2022} who involved as well delayed exponential. 
\end{table*}

    In summary, at $z<1$ the SPS (CB19 versus BC03 versus M05) dominates the systematic offsets of our configurations, while at $z=1.0$--1.5 the code/SFH choices contribute also significantly. It is evident that configuration choices can  introduce systematic shifts typically of $\sim$0.5~dex, which we will see in the next section are comparable to the scatter among different GSMF studies from different authors at high redshift.

\begin{figure*}[t]
    \centering
    \includegraphics[width=\textwidth]{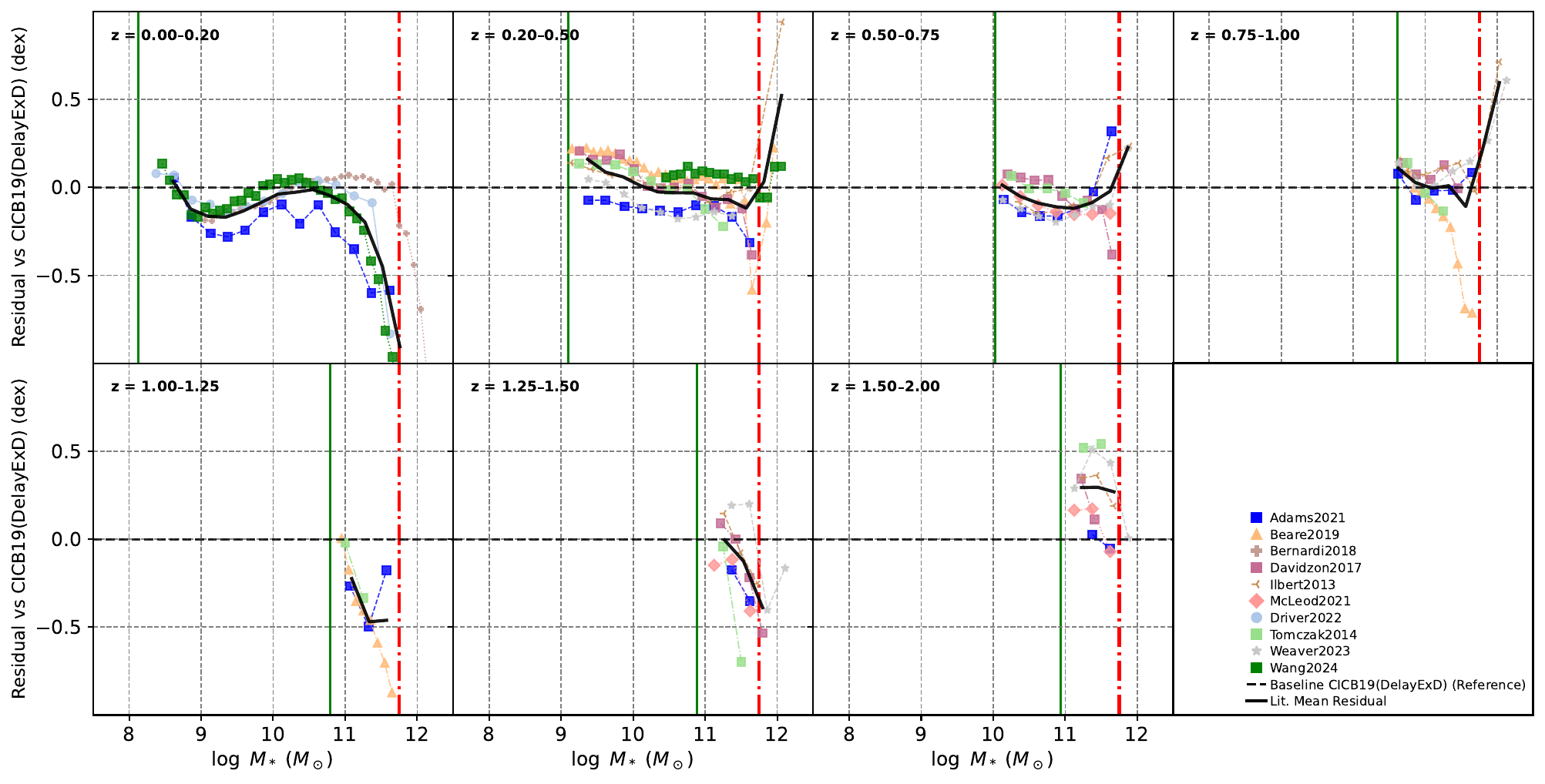}
    \caption{Residuals of literature stellar mass functions with respect to the KiDS DR4 four-model mean, shown across redshift bins. This comparison highlights the level of systematic scatter among different observational datasets. The symbols are the same as Figure~\ref{fig:smf_results}, the black solid curve is the mean of residual of literature compiled in each redshift bin, and will keep the same in the following figures.}
    \label{fig:smf_literature_residuals}
\end{figure*}

    \subsection{GSMF results compared with previous literature}
    \label{sec:smf_lits}

    In this section we discuss the evolution of the GSMF from KiDS DR4 and how it compares with previous studies. Most surveys generally follow two trends: either they achieve high redshift coverage (due to deep magnitude limits) over relatively small survey areas, making them susceptible to cosmic variance and potentially missing information for the high-mass end, or they cover large areas but are limited to lower redshifts (due to shallower magnitude limits). Our work attempts to bridge this gap by constructing a GSMF at relatively high redshifts over a large survey area, based on the KiDS $u$-to-$K_s$ bands (i.e. spanning optical and near-infrared wavelengths).

    Our literature compilation encompasses surveys with great  diversity in observational depth and coverage. As seen in table \ref{tab:literature} the deepest observations include the COSMOS/XMM-LSS survey reaching $u^*=27.2$ mag \citep{Adams2021} and COSMOS2020 achieving $F814W\sim28$ mag \citep{Weaver2023}, while the largest area coverage is provided by SDSS DR7 spanning 4700 deg$^2$ \citep{Bernardi2018} and GAMA covering 230 deg$^2$ \citep{Driver2022} and DESI SV3 covering 133 deg$^2$ \citep{Wang2024}. The range of volumes involves data from ultra-deep pencil-beam surveys ($\sim$0.1 deg$^2$) to wide-area surveys ($>$1000 deg$^2$), providing complementary constraints across different stellar masses and different redshift regimes. We note that compared with previous studies, our survey volume is typically much higher and can reach 1 $Gpc^3$ up to z=1 and 10 $Gpc^3$ up to z=2.

    Redshift determination in most studies employs two primary methodologies: spectroscopic redshifts offering the highest precision (SDSS DR7, GAMA, DESI) typically available at $z \sim 0$, and photometric redshifts derived through SED fitting techniques for $z > 0.2$. The photometric redshift estimates predominantly utilize template-based codes such as LePhare \citep{Arnouts1999,Ilbert2006} and EAZY \citep{Brammer2008}, incorporating stellar population synthesis models from \citet{Bruzual2003} and extensive template libraries covering diverse galaxy types from star-forming to quiescent populations. We remind that our redshift determination is done using a deep learning based method which has some advantages with respect traditional photometric redshift estimations when 9-bands photometry is considered (Figure~\ref{fig:z-comparison} and \citet{Li2022,Xielh2023}).

    Regarding the stellar mass estimation across the studies considered, consistently relies mostly on SED fitting methodologies (except \citet{Beare2019}, who used $K$-band stellar mass-to-light ratio conversions). Most authors typically use \citet[][BC03]{Bruzual2003} stellar population synthesis models and \citet{Chabrier2003} IMF. The fitting procedures usually assume exponentially declining star formation histories and incorporate dust attenuation following \citet{Calzetti2000} prescriptions. 

 For clarity, we show the residuals of each individual work with respect to our KiDS reference that adopts an updated CB19 SPS in Figure~\ref{fig:smf_literature_residuals}, highlighting that indeed the GSMF can deviate from author to author. The discrepancies within the literature (e.g. \citet{McLeod2021} and \citet{Driver2022} at z = 0 , or \citet{Adams2021} and \citet{Beare2019} at z = 1) are due to the fact that different studies  adopt different SED-fitting setups, cover different volumes and have different wavelength coverage.  Differences are evident  at the level of several tenths of a dex (typically around 0.5 dex) but larger disparities are also common, like for example \citet{Wang2024} vs \citet{Ilbert2013} – 0.6 dex at z = 0.2-0.5 and \citet{Adams2021} vs \citet{Davidzon2017}-0.6 dex at z = 0.5-0.75 and \citet{Weaver2023} vs \citet{Tomczak2014} - 1 dex at 1.25-1.50.

    According to Figure \ref{fig:smf_literature_residuals}, at low redshifts ($z<0.2$), our results show good agreement with previous literature (considering the 0.5 dex uncertainty among different studies), showing values between all different authors down to $10^{8.5} M_\odot$. At $\log(M_*\ /M_\odot) \le 8.5$, near the mass limit line, our results show an offset of about 0.1--0.3 dex lower. This could be due to some limitations of GaZNet-z, which in the future it will be updated to deeper training spectroscopic redshift dateset. For \citet{Wang2024} at $z\approx 0$, their high-mass number densities are lower than ours, which is not unexpected given their lower magnitude selection; these differences can be amplified at the extreme massive end where completeness and Eddington scatter are critical. Our results for the high mass end are also typically higher than those of \citet{Adams2021}. We note that for z = 0 our GSMF for high masses  ($>10^{11.5} $ $M_\odot)$ has higher values than most studies.

    We note that for galaxies at the lowest redshifts (z = 0.001--0.200), our measured masses near the low mass limit are typically lower than those in the literature and this  suggests that our adopted method underestimates the true mass limit in this low redshift bin. We therefore adopt a more conservative limit of $\log(M_*/M_{\odot}) \simeq 8.5-9.0$ at z = 0.001--0.200 for the remainder of this paper, acknowledging limitations of our survey for low mass objects at z $\sim$ 0. 

    At z = 0.2 to 1.0, we note that our reference GSMF has excellent agreement with most previous literature measurements up to the limit of $10^{11.75} $ $M_\odot$ \footnote{The red vertical line describes the $10^{11.75} $ $M_\odot$ mass that represents the upper limit that can be reached by most studies.} However, for objects with extremely high masses of $> 10^{11.75} $ $M_\odot$ the results of \citet{Ilbert2013} and \citet{Weaver2023} are typically implying higher values by $0.5$ dex. The uncertainties reported from  \citet{Ilbert2013} for their high mass bins are quite large ($\sim 0.5$ dex) so the disparities with our reference  can be attributed to differences in SED fitting and cosmic variance. In addition, the values reported by  \citet{Weaver2023} for objects with masses of $> 10^{11.75}$ $M_\odot$ are considered upper limits by the authors. Regardless, we find a considerable number of high mass objects in our analysis too, even if our SMFs are  relatively lower.
    
    Tracing the trend  to higher redshifts, we find that at z = 1.00–1.25, our reference GSMF is higher than the literature average by 0.35 dex. Moving to z = 1.25–1.50, once again our values are slightly larger by 0.2 dex  than those in the literature, with the difference concentrated at the high-mass end. At the higher redshift range of z = 1.5–2.0, the situation reverses, with our average falling slightly by 0.25 dex below the literature average. 
    
    Overall, an interesting aspect of the large KiDS DR4 area is that it allows us to probe galaxies at the very high-mass tail. We find a non-negligible population of galaxies with $\log(M_*/M_\odot)>11.75$, whose number densities appear to follow a steep power-law extension instead of an exponential cut-off implied from a Schechter form. This is in agreement with the recent studies of \citet{Rodriguez-Puebla2020} and \citet{Vazquez-Mata2025} who also  support a potential excess at the high-mass end. 
    
    \citet{Vazquez-Mata2025} derived their SMF using data from the MaNGA Visual Morphology (MVM) catalogue at $z \sim 0$, separating it by morphological type. The authors found that a triple Schechter function provides a better description of the total GSMF than a single Schechter form, associating three characteristic masses with different galaxy types. In agreement with our work, the authors also report an excess of massive objects relative to the predictions of a single Schechter function.

    \citet{Rodriguez-Puebla2020} determined the galaxy SMF for a range of $M_* \sim 3 \times 10^7$ to $3 \times 10^{12}\,M_\odot$ at $z = 0$, by combining two spectroscopic samples from the Sloan Digital Sky Survey in the redshift range $0.0033 < z < 0.20$. Stellar masses were derived using five colour-dependent mass-to-light ratios that had  differences among them of approximately 0.5–1 dex coming from  uncertainties in  SPS modeling. As their fiducial definition, the authors adopted the geometric mean of these five stellar masses per galaxy. The SMF is well described by a function combining a sub-exponential Schechter function and a broken power law since there is an excess of massive objects relative to what would be expected from a single Schechter function, once again in agreement with our results.  
    \begin{figure*}[t]
    \centering
    \includegraphics[width=0.95\textwidth]{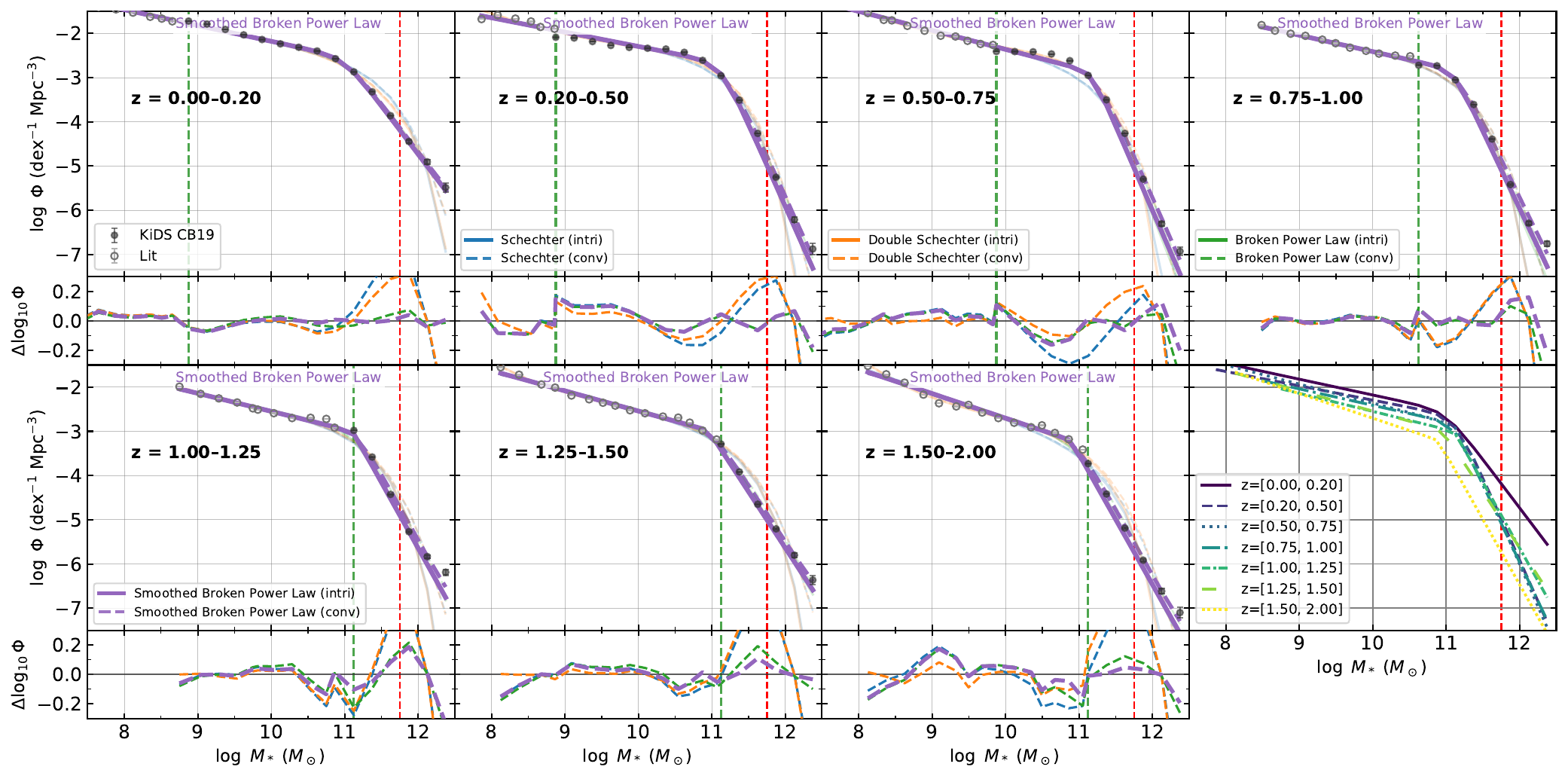}
    \caption{Stellar mass functions with Eddington bias correction using direct fitting. Different models are compared using the Gaussian-Lorentzian kernel choice following \citet{Weaver2023}.}
    \label{fig:smf_eb_weaver_cutmass}
\end{figure*}
    We stress though that measuring the high mass end of the GSMF is particularly challenging and our analysis suggests that systematic uncertainties and limitations in derived redshifts have to be thoroughly investigated. Future datasets with even larger volume covered, with updated SED fitting strategies and redshift estimation techniques will be required to confirm this behavior, as we heed caution that the uncertainties related to SPS, SFHs, code and redshift determination are important. In the following section we demonstrate that the result of excess objects at the high mass end with respect a single Schechter  does not change upon Eddington bias correction.

\subsection{Eddington bias correction and analytical formulas}
\label{sec:eddington_bias}

We correct for Eddington bias, which arises from the uncertainty-driven redistribution of galaxies across mass bins of the SMF and to get the intrinsic GSMFs. We express stellar mass as $x\equiv\log_{10}(M_*/M_\odot)$. The intrinsic true mass function $\Phi_{\rm true}(x)$ is related to the observed mass function $\Phi_{\rm obs}(x)$ through a convolution with an uncertainty kernel $K(x_{\rm obs}|x_{\rm true})$:
\begin{equation}
    \Phi_{\rm obs}(x_{\rm obs}) = \int K(x_{\rm obs}|x_{\rm true}) \Phi_{\rm true}(x_{\rm true}) dx_{\rm true}
\end{equation}

To model the intrinsic mass function $\Phi_{\rm true}$, we consider several analytical forms. For clarity, we write the functional forms below in linear $\Phi$ units as functions of $x=\log_{10}(M_*/M_\odot)$, while the actual fitting is performed in $\log_{10}\Phi$ space. We set $\Delta x \equiv x-x^*$ (or $x-x_{\rm ref}$/$x-x_{\rm break}$ depending on the model). The functional forms are:

\begin{itemize}
    \item Schechter function:
    \begin{equation*}
        \Phi_{\rm Sch}(x) = \ln(10)\,\Phi_*\,10^{(\alpha+1)\Delta x}\exp\left(-10^{\Delta x}\right)
    \end{equation*}
    
    \item Double Schechter function:
    \begin{equation*}
        \Phi_{\rm dSch}(x) = \ln(10)\,\exp\left(-10^{\Delta x}\right)\sum_{i=1}^{2}\Phi_i\,10^{(\alpha_i+1)\Delta x}
    \end{equation*}
    
    \item Broken power-law function (Double power-law):
    \begin{equation*}
        \Phi_{\rm dPL}(x) = \frac{\Phi_0}{10^{-\alpha\Delta x}+10^{-\beta\Delta x}}
    \end{equation*}
    
    \item Smoothed Broken Power-Law function: 
    A broken power-law with a smooth transition controlled by a smoothness parameter $s$:
    \begin{equation*}
        \Phi_{\rm smoothed}(x) = \Phi_0\,10^{\alpha\Delta x}\left(1+10^{\Delta x/s}\right)^{-s(\alpha-\beta)}
    \end{equation*}
    where $\Delta x = x-x_{\rm break}$, $\alpha$ and $\beta$ are the slopes at the low- and high-mass ends respectively, and $s$ controls the width of the transition region.
\end{itemize}
\vspace{0.5em}
\noindent The factor $\ln(10)$ appears in the Schechter-type functions because the GSMF is expressed per dex in $x=\log_{10}(M_*/M_\odot)$.
\begin{figure*}[t]
    \centering
    \includegraphics[width=0.95\textwidth]{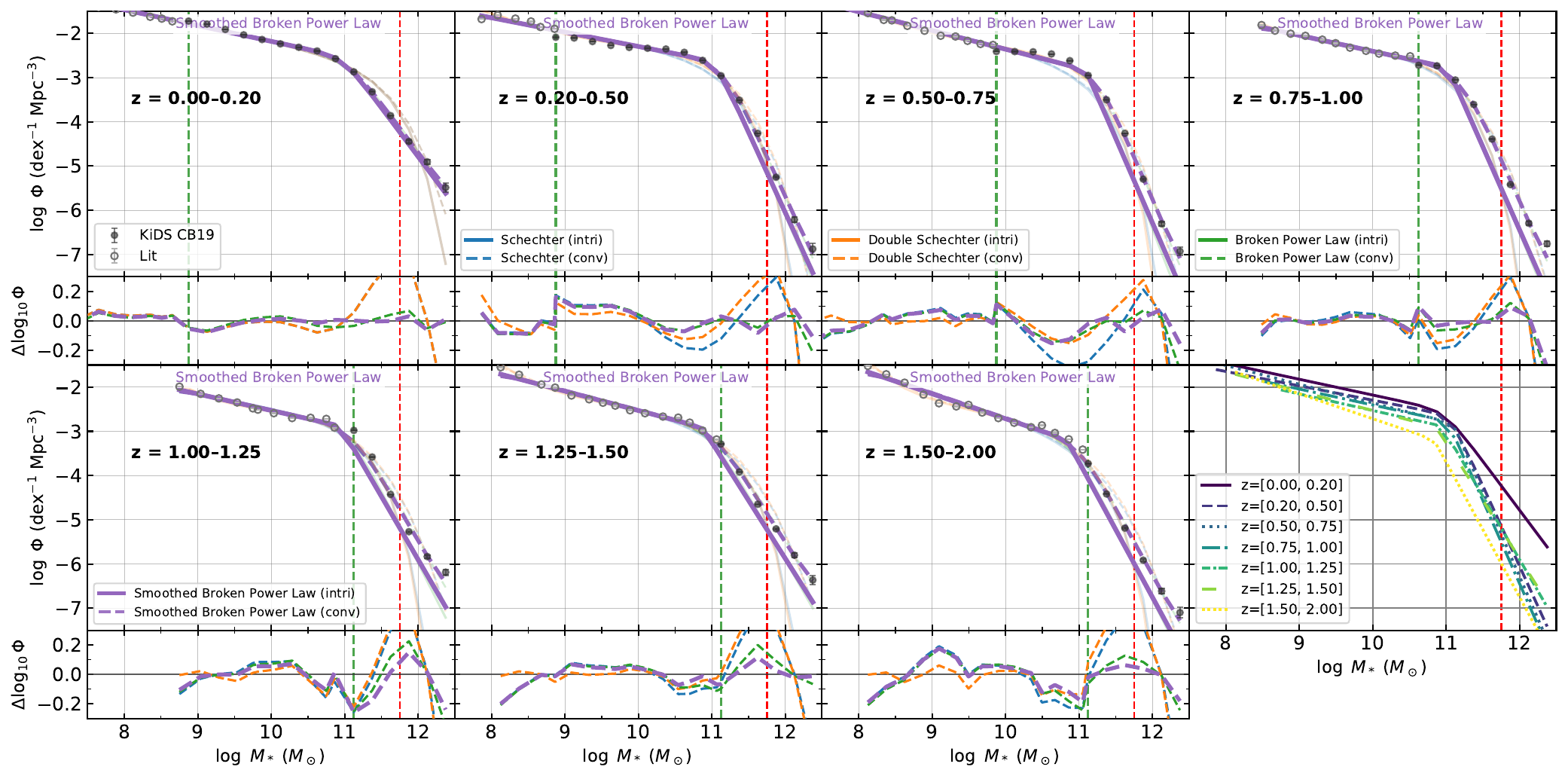}
    \caption{Stellar mass functions with Eddington bias correction using direct fitting method. Different models are compared using the redshift-dependent Gaussian kernel approach following \citet{Chaikin2025}.}
    \label{fig:smf_eb_chaikin_nocutmass}
\end{figure*}
To recover the intrinsic parameters of the SMF, we perform a direct least-squares fit to the observed data, combining our results at high mass end and literature low mass end according to out mass completeness limits in each redshift bin. Although the analytical forms above are shown in linear $\Phi$ units to make their structure clear, the fitting minimizes the residuals between the convolved model predictions and the observed SMF measurements in $\log_{10}\Phi$ space. While MCMC sampling is available for full uncertainty quantification, the results presented here rely on this direct fitting approach for its efficiency and stability. The procedure involves: (1) defining the analytical intrinsic model (as above); (2) constructing the redshift-dependent uncertainty kernel; (3) convolving the model with each kernel choice; (4) optimizing the parameters to match the data; and (5) finally obtaining the parameters of each intrinsic model. For step (3), we use the mean absolute difference (MAD) to quantify the overall residual amplitude between each kernel-convolved model and the observational data:
\begin{equation}
    {\rm MAD} = \frac{1}{N}\sum_{i=1}^{N}\left|\log_{10}\Phi_{\rm mod, i}-\log_{10}\Phi_{\rm obs, i}\right|,
    \label{eq:mad}
\end{equation}
where $\Phi_{\rm mod}$ is the kernel-convolved model GSMF, $\Phi_{\rm obs}$ is the observed GSMF, and $N$ is the number of stellar-mass bins used in the comparison. This framework allows us to robustly correct for Eddington bias while well describing the observed high-mass-end behavior.

\subsubsection{Correction using Gaussian-Lorentzian kernel (\citet{Weaver2023})}
\label{subsec:kernel_weaver}

Following \citet{Weaver2023}, we adopt a Gaussian-Lorentzian mixture kernel choice ($K_{\rm gauss-lorentz}$) to account for non-Gaussian tails in mass uncertainties, which are particularly significant at higher redshifts. Using the notation of the previous section, the kernel is defined as:
\begin{equation}
    K_{\rm gauss-lorentz}(\Delta x) \propto \exp\left(-\frac{(\Delta x)^2}{2\sigma^2}\right) \cdot \frac{\tau^2}{\tau^2 + (\Delta x)^2}
\end{equation}
where $\Delta x = x_{\rm obs} - x_{\rm true}$, $\sigma = 0.2$ is the Gaussian width, and $\tau$ controls the Lorentzian tail. The Lorentzian scale evolves with redshift as $\tau(z) = \tau_0 (1 + z)$, with $\tau_0 = 0.1$. This kernel choice is expected to effectively capture the uncertainty-driven redistribution of massive galaxies across mass bins due to large photometric errors.
 
 We present the effect of correcting the SMF using the above kernel choice in Figure~\ref{fig:smf_eb_weaver_cutmass}, while the fitted parameters for the different forms can be found in the Supplementary Materials. The best performing functions are typically the smoothed broken power law and the broken power law, besides the fact that the parameter estimation can come with large uncertainties. A single Schechter form is unable to capture the high mass end bins. We find that the EB corrected SMF has slightly lower values with respect the uncorrected at z = 0.00-0.75. At higher redshifts the SMF at the high mass end can be lower by $\sim  0.2-0.4 $ dex. We also present the fitting results with $\log M_*/M_{\odot}<11.75$ in the Supplementary Materials, where the double Schechter function demonstrates best performance.

\begin{figure*}[t]
    \centering
    \includegraphics[width=0.95\textwidth]{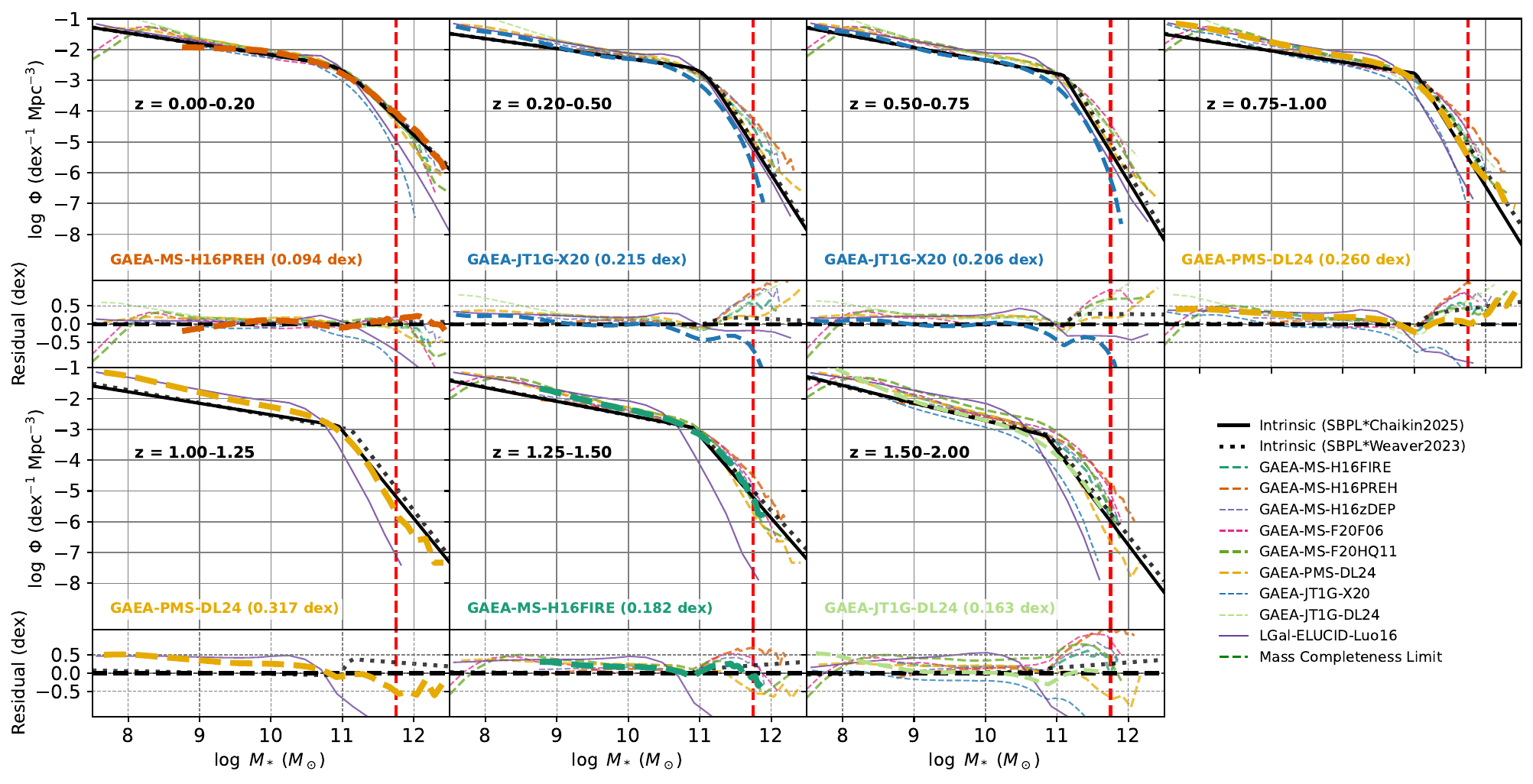}
    \caption{Comparison of stellar mass functions (SMFs) across redshift bins (z = 0.0--2.0) between KiDS observations and various SAM models, after applying an Eddington-bias correction to the observed data. We present residual plots for clarity.}
    \label{fig:smf_sam_comparison_eb}
\end{figure*}
\subsubsection{Correction using Redshift-dependent Gaussian kernel (\citet{Chaikin2025})}
\label{subsec:kernel_chaikin}

Alternatively, following \citet{Chaikin2025}, we implement a simpler Gaussian kernel choice with a redshift-dependent width. This kernel choice assumes that the primary source of uncertainty growth is the decreasing signal-to-noise ratio in photometric data at higher $z$. The kernel is defined as:
\begin{equation}
    K_{\rm gaussian}(\Delta x) = \frac{1}{\sqrt{2\pi}\sigma(z)} \exp\left(-\frac{(\Delta x)^2}{2\sigma(z)^2}\right)
\end{equation}
where the width evolves as:
\begin{equation}
    \sigma(z) = \min\left(0.1 \cdot (1 + z), 0.3\right)
\end{equation}
The upper limit of $0.3$ dex prevents the kernel from becoming excessively broad, reflecting the saturation of uncertainty growth in deep survey fields.

We present the effect of correcting the SMF using the above kernel choice in Figure~\ref{fig:smf_eb_chaikin_nocutmass}, while the fitted parameters for the different forms can also be found in the Supplementary Materials. The best performing functions are once again the smoothed broken power law and the broken power law. We once again find that the EB corrected SMF has slightly lower values with respect the uncorrected at z = 0.00-0.75. However, at higher redshifts the SMF at the high mass end can be lower by $\sim  0.5 $ dex for objects with $\log(M_*/M_\odot)\lesssim 11.75$ but even 1 dex for higher masses. The resulting characteristic mass $\log_{10}M_*^*$ is consistent with previous works within $1\sigma$. Compared to the \citet{Weaver2023} kernel choice, the \citet{Chaikin2025} kernel choice yields larger high-mass corrections because its redshift-dependent Gaussian width can become broader at high redshift.

    \subsection{GSMF results compared with theoretical models}
    \label{sec:discussion}

    In this section, we compare our EB corrected GSMFs (via the method discussed in subsection \ref{subsec:kernel_chaikin}) with the predictions from semi-analytic models (SAMs) and cosmological hydrodynamical simulations. We examine multiple SAM variants from the L-Galaxies and GAEA families, as well as several widely used hydrodynamical simulations (IllustrisTNG, EAGLE, and Simba). We also present a new combination of the Jiutian simulation \citep{Han2025} with  the updated GAEA model from \citet{DeLucia2024}. We refer to this new configuration as \textit{GAEA-JT1G-DL24} in the following content. These models span diverse physical prescriptions for star formation, stellar and AGN feedback, and black hole growth, with varying calibration strategies ranging from single-epoch ($z\approx 0$) to multi-epoch ($z=0$--3) approaches. A comprehensive summary of the physical ingredients and calibration tactics for each model, including specific variants and references, is provided in the Supplementary Materials.

    Since most models have actually been  calibrated  against previous observational GSMFs (typically at $z\approx 0$, and for some models even up to $z\sim 3$),  our comparison is  mostly intended to be qualitative. However, an other important aspect to keep in mind is that typically the calibration of the parameters of these models (related to feedback and star formation) was usually done against observations that are impacted by Eddington bias. Thus, upon considering the effects EB (and the improvements in volume, updated SPS and redshift determination) it is not guaranteed that models will show good performance.
    To summarize the level of agreement between observations and models in a uniform/transperent way, we also use the  residuals between each model GSMF and our EB corrected GSMFs in each redshift bin. The GSMF comparison figures are ideal to  study differences in shape, whereas the residual figures are better at capturing more accurately the  normalization offsets and mass-dependent mismatches. We then calculate the mean absolute difference (MAD; Eq.~\ref{eq:mad}) cross stellar mass range above our mass limit to illustrate how each model performs overall in a redshift bin. 

    For the semi-analytic models we show the model--data comparison in  Figure~\ref{fig:smf_sam_comparison_eb}. We note that not all model families provide snapshots at every redshift bin shown in our figures. Within our SAM set, the configuration yielding the lowest MAD (i.e. having the "best" comparison with our mean/referece observations) varies across redshift bins: GAEA-MS-H16FIRE shows lowest MAD at $z<0.2$,GAEA-JT1G-X20 at $z=0.2$--0.75, GAEA-PMS-DL24 at $z=0.75$--1.25 and $z=1.5$--2.0, while GAEA-MS-H16FIRE at $z=1.25$--1.5. At highest redshifts GAEA-JT1G-X20 performs quite well. We also note that these specific models showed better performance upon considering the effects of EB. In general, while modern SAMs broadly succeed in reproducing the {\it shape} of the GSMF  (Figure \ref{fig:smf_sam_comparison_eb}), non-negligible quantitative tensions according to the residual analysis persist with respect the EB corrected observations, especially at the high mass end that can exceed 0.5 dex.

\begin{figure}[t]
    \centering
    \includegraphics[width=0.95\textwidth]{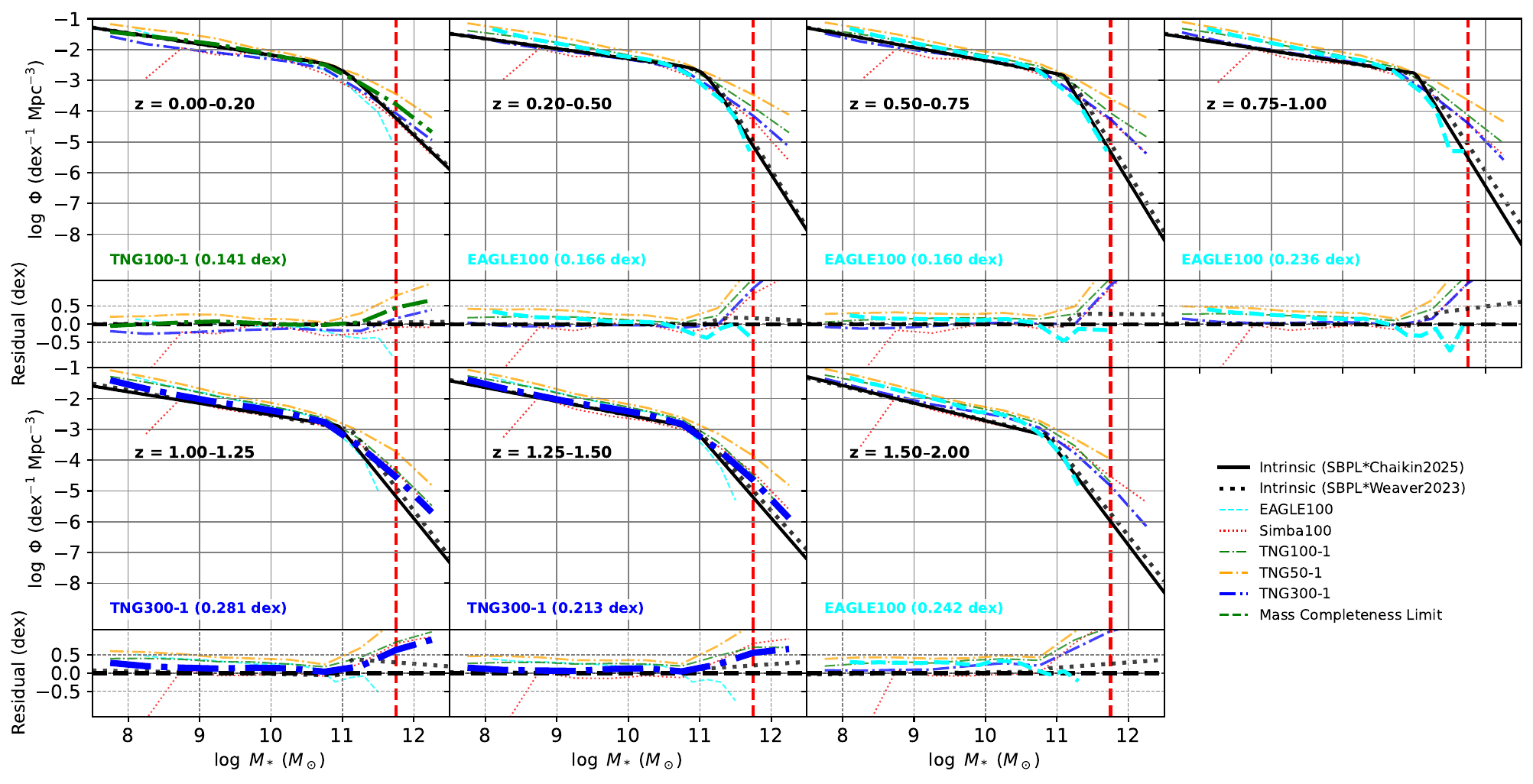}
    \caption{Comparison of stellar mass functions (SMFs) across redshift bins (z = 0.0--2.0) between KiDS EB corrected observations and various hydrodynamical simulations. Alongside we present the mean absolute differences (MAD). The EAGLE simulations demonstrate the lowest residuals with respect observations.}
    \label{fig:smf_hydro_comparison_eb}
\end{figure}

     In  Figure~\ref{fig:smf_hydro_comparison_eb} we perform the same comparison as \ref{fig:smf_sam_comparison_eb} but this time for cosmological hydrodynamic simulations. Using the MAD metric defined in Eq.~\ref{eq:mad}, we find that EAGLE yields comparatively the smallest offsets with respect the EB corrected observations in most redshift bins ($z=0.2$--1.0, and $z=1.5$--2.0), while TNG100-1 shows lowest MAD at $z=0$--0.2 and TNG300-1 at $z=1.0$--1.5. 
     
     However, we have to note that some models at the high mass end significantly over-predict galaxies with respect our EB corrected observations. In general, at most redshifts, the model--data offsets remain modest in the low/intemediate-mass regime, where most models show residual$\lesssim 0.5$~dex.  At higher masses ($\log(M_*/M_\odot)>10.75$, EAGLE100 show persistently minus residual across $z<2$, while TNG and Simba can show deviations with respect observations even by +0.9 dex.
\end{multicols}

\begin{multicols}{2}

     In our work we determine the reference observed GSMFs by incorporating updated SPS templates \citep{CB19}, we use a survey with large area and exploit deep learning methodology for redshift determination. 
     The comparison with some simulations (after accounting for Eddington bias) reveals large inconsistencies well beyond the scatter seen in observational studies (Figure \ref{fig:smf_literature_residuals}). We demonstrate that the effect of EB can play a crucial role for objects in the high mass end. We therefore suggest, that in order to make an optimal tuning procedure and a fair comparison against observations, future model calibrations should be done against Eddington-bias-corrected GSMFs.

    \section{Summary and Conclusions}
    \label{sec:summary}

    In this work, we have presented a comprehensive analysis of the GSMF using the Kilo-Degree Survey Data Release 4 (KiDS DR4), covering an effective area of 676.9 deg$^2$ after quality cuts. To study systematic uncertainties, we combined two SED codes (CIGALE, LePhare), and two different SFHs (delayed exponential declining and exponential declining, respectively) with three stellar population models (CB19, BC03, M05). We adopted the CIGALE+CB19(CICB19) as our our fiducial measurement.
    Our study leverages deep multi-band photometry (nine bands from $u$ to $K_s$) combined with sophisticated  machine learning photometric redshift estimation (GaZNet-z) and multiple SED fitting approaches to measure the GSMF across the redshift range $0.001 < z < 2.0$. Our main results are summarized as follows: 

\begin{itemize}
\item The large survey area of KiDS provides excellent statistics for studying the high-mass end of the GSMF, complementing  other efforts done by deeper but smaller-area surveys. Our GSMF measurements show an overall consistency with previous literature within the mass-complete regime, especially for objects with $\log(M_*/M_\odot)\lesssim 11.75$, considering systematics. 

\item The differences demonstrated in our work between different observational studies underscore the importance of understanding survey-specific effects (e.g. survey area), and selection effects (e.g. magnitude selection) in GSMF measurements. In addition, in our work by using different configurations for the SED fitting we demonstrate that offsets of 0.6 dex can also be easily attributed to uncertainties/differences in assumptions in the SPS modeling and SFH.

\item  Our results for objects with $\log(M_*/M_\odot)>11.75$, suggest a less steep high-mass end extension and not a purely exponential Schechter like cut-off (i.e. there are more massive galaxies than what it would be expected from a simple Schechter form), even when a considerable fraction of objects is excluded due to redshift determination uncertainties.  At $z\sim 0.2$--1.0, the ``extreme high mass end" ($\log(M_*/M_\odot)>12$) shows a tilting behavior. However, we draw caution in interpreting this result as we have already shown that GSMFs can be subject to systematics (e.g., redshift/SED-fitting uncertainties and/or mild non-galaxy contamination).

\item We present analytical forms (schechter functions, double schechter functions, broken power laws) to describe our results. We find that typically double schechter and broken power laws perform better to describe the observed GSMF. In addition, we investigate two different ways to correct our observations for Eddingthon bias effects. The correction results in lower values for the GSMF by 0.1 dex for the high mass end.

\item We compare our EB corrected measurements  and the literature compilation with multiple semi-analytic models (SAMs) from the L-Galaxies and GAEA families, as well as several widely used hydrodynamical simulations (EAGLE, IllustrisTNG, Simba). We quantify model--data agreement using mean absolute difference (MAD) of residuals as a diagnostic metric. We provide comparisons between our reference GSMFs. Typically simulations show higher values for the GSMF, especially at the high mass end, that can exceed 0.5 dex.

\item We note that usually the comparisons between models and observations are done in the literature without explicitly accounting for observational stellar-mass scatter and the resulting Eddington bias.  It is evident that when EB effects are considered, typically most models demonstrate even worse comparison with respect observations, enhanced typically by an additional 0.2-0.4 dex. However, this is not the case for all simulations (e.g. EAGLE100) that actually shows better comparison with observations once EB effects are considered. Among SAMs, the lowest-MAD configuration varies across redshift bins: GAEA-family models (e.g. GAEA-PMS-DL24) shows lowest MAD at $z<1.5$. The particularly high MADs for high mass objects for some models, possibly reflects limitations for the uncertain feedback prescriptions employed.

\end{itemize}

    Looking forward, future model calibrations should either tune against Eddington-bias-corrected GSMFs or forward-model the observational scatter when comparing to data, to avoid systematic biases in parameter constraints. In addition, mock catalogues \citep{Trayford2015,katsianis2020,Baes2020,Yizhou2024} can be a valuable tools to investigate how different assumptions in SED fitting can affect the derived GSMFs. Future studies will benefit from combining optical surveys like KiDS with deep near- and mid-infrared observations to minimize systematic uncertainties in stellar mass estimation. Considerable efforts have to be done in estimating Photometric and spectroscopic redshifts. The Euclid mission and the Vera C. Rubin Observatory's Legacy Survey of Space and Time (LSST) will provide unprecedented combinations of survey area and depth, enabling more precise GSMF measurements across cosmic time. The methodological framework developed in this study, particularly the explicit treatment of Eddington bias,  and the use of multiple SED fitting configurations to clarify systematics, offers/suggests a practical approach for analyzing such datasets.
    \Acknowledgements{Q.W. and A.K has been supported by the National SKA Program of China (2025SKA0150104), the 100 talent program of the Sun Yat-sen University and the Guangdong Basic and Applied Basic Research Foundation with No 2025A1515012670. J.H acknowledges support from NSFC 12595312, National Key R \& D Program of China (2023YFA1607800, 2023YFA1607801), and China Manned Space Program (CMS-CSST-2025-A04, CMS-CSST-2021-A03).}

    \InterestConflict{The authors declare that they have no conflict of interest.}

    \section*{Acknowledgements}
    Q.W. and A.K has been supported by the National SKA Program of China (2025SKA0150104), the 100 talent program of the Sun Yat-sen University and the Guangdong Basic and Applied Basic Research Foundation with No 2025A1515012670. J.H acknowledges support from NSFC 12595312, National Key R \& D Program of China (2023YFA1607800, 2023YFA1607801), and China Manned Space Program (CMS-CSST-2025-A04, CMS-CSST-2021-A03).
    
    \InterestConflict{The authors declare that they have no conflict of interest.}

    \section*{Conflict of interest}
    The authors declare that they have no conflict of interest.

    \bibliographystyle{Configurations/scpma-custom}
    \bibliography{Configurations/Brief_KiDSDR4SMF}

\end{multicols}

\clearpage

\end{document}


\raggedcolumns

\ensubject{Supplementary Materials}
\ArticleType{Supplementary Materials}
\SpecialTopic{}
\Year{2026} \Month{June} \Vol{66} \No{1} \DOI{??} \ArtNo{000000} \ReceiveDate{March 30, 2026}
\AcceptDate{??}

\title{Supplementary Materials for ``Growth from KiDS: Stellar mass assembly of galaxies since z=2 in the light of Kilo-Degree Survey''}
{Supplementary Materials for KiDS DR4 GSMF}

\AuthorCitation{Wang Q, Katsianis A, Napolitano N R, et al.}

\author[1]{Qingshan Wang}{}
\author[1]{Antonios Katsianis}{katsianis@sysu.edu.cn}
\author[2,3]{Nicola R. Napolitano}{}
\author[1]{Linghua Xie}{}
\author[4]{Rui Li}{}
\author[5,6,7]{Xiaohu Yang}{}
\author[1]{\\Baitian Tang}{}
\author[5,6]{Jiaxin Han}{}
\author[5,6]{Zhenlin Tan}{}
\author[8]{Lizhi Xie}{}
\author[1]{Ying Tang}{}
\author[1]{Yuchang Li}{}
\author[1]{Hangxin Pu}{}

\address[1]{School of Physics and Astronomy, Sun Yat-sen University, Zhuhai Campus, 2 Daxue Road, Xiangzhou District, Zhuhai, 519082, China}
\address[2]{Department of Physics E. Pancini, University Federico II, Via Cinthia 21, I-80126 Naples, Italy}
\address[3]{INAF -- Osservatorio Astronomico di Capodimonte, Salita Moiariello 16, I-80131 Napoli, Italy}
\address[4]{Institute for Astrophysics, School of Physics, Zhengzhou University, Zhengzhou, 450001, China}
\address[5]{Department of Astronomy, Shanghai Jiao Tong University, Shanghai 200240, China}
\address[6]{State Key Laboratory of Dark Matter Physics, Key Laboratory for Particle Astrophysics and Cosmology (MOE), \\ \& Shanghai Key Laboratory for Particle Physics and Cosmology, Shanghai Jiao Tong University, Shanghai 200240, China}
\address[7]{Tsung-Dao Lee Institute and State Key Laboratory of Dark Matter Physics, Shanghai Jiao Tong University, Shanghai 202110, China}
\address[8]{Astrophysics Center, Tianjin Normal University, Tianjin 300387, China}

\abstract{This supplementary document provides supporting material for the KiDS DR4 galaxy stellar mass function analysis, including details of the galaxy-formation models, Eddington-bias tests with a high-mass cut, GSMF data tables, best-fit analytical parameters, and a summary of the KiDSDR4-C0v1.1 catalog columns.}

\keywords{galaxies: formation; galaxies: evolution; galaxies: statistics; galaxies: stellar content; surveys}
\PACS{47.55.nb, 47.20.Ky, 47.11.Fg}
\maketitle

\setcounter{section}{0}
\renewcommand{\thesection}{S\arabic{section}}
\renewcommand{\thetable}{S\arabic{table}}
\renewcommand{\thefigure}{S\arabic{figure}}

    \begin{multicols}{2}

        \section{Galaxy Formation Models: Physical Prescriptions and Calibration Strategies}
        \label{sec:appendix_models}

        The galaxy formation models employed in this work encompass a wide range of physical prescriptions and calibration strategies. Understanding the diversity {\it and uncertainty}  of these approaches is essential for interpreting the systematic differences in model predictions presented in the main text. Table~\ref{tab:model_comparison} provides a comprehensive summary of the key physical ingredients across all models examined in this study, including their star formation laws, stellar and AGN feedback implementations, black hole growth prescriptions, and calibration strategies. In this section, we provide an overview of these physical modules, categorize the various methodological approaches, and discuss the implications of different calibration strategies.

        \subsection{Classification of Physical Prescriptions}

        The physical modules implemented in galaxy formation models can be broadly categorized into several key components, each with distinct methodological variants across the model suite we examine.

        \textit{Star Formation Laws.} The models adopt fundamentally different prescriptions for converting cold gas into stars. The simplest approach, used in models, employs a threshold-based conversion that scales with dynamical time, where star formation occurs only when cold gas exceeds a critical surface density. More sophisticated treatments explicitly partition atomic and molecular hydrogen, with star formation tied exclusively to molecular gas surface density following a Schmidt law. This H$_2$-based approach is adopted by the GAEA family (including LGal-ELUCID-Luo16, GAEA-MS-H16FIRE, GAEA-MS-F20F06, GAEA-PMS-DL24, GAEA-JT1G-X20, and GAEA-JT1G-DL24), as well as Simba100. Hydrodynamical simulations typically implement density-dependent star formation with varying prescriptions: EAGLE uses a pressure-law formulation where star formation rate scales as $P^{1.5}$, while the IllustrisTNG suite adopts an effective equation of state with density thresholds around $0.1\,\mathrm{cm}^{-3}$ for TNG100/300 and higher thresholds for TNG50.

        \textit{Stellar Feedback Mechanisms.} Stellar feedback implementations vary significantly in their approach to energy injection and gas redistribution. Ejective feedback schemes, adopted by most SAMs and several simulations, reheat or eject cold gas through supernova-driven winds, with mass-loading factors that typically scale with halo virial velocity or dark matter velocity dispersion. The GAEA-MS-H16FIRE variant implements the strongest ejective feedback by adopting FIRE-calibrated mass loading with explicit redshift dependence $(1+z)^{1.25}$. In contrast, GAEA-MS-H16PREH explores a preventative feedback approach that reduces initial gas accretion in low-mass halos rather than expelling gas after it has cooled. Hydrodynamical simulations implement feedback through various schemes: EAGLE use stochastic thermal heating where gas particles are heated to specific temperatures ($10^{7.5}$ K for EAGLE stellar feedback), while IllustrisTNG employs isotropic kinetic winds with velocities scaling with dark matter velocity dispersion and metallicity-dependent energy. The reincorporation of ejected material follows different timescales across models, generally inversely proportional to halo mass to delay gas return in lower-mass systems-a key ingredient for reproducing downsizing trends.

        \textit{Black Hole Growth Prescriptions.} Models implement black hole accretion through diverse physical channels. Traditional SAM approaches combine merger-driven cold gas accretion (quasar mode) with hot halo accretion (radio mode), as seen in the L-Galaxies and early GAEA variants. More recent SAMs like GAEA-MS-F20F06, GAEA-MS-F20HQ11, and GAEA-PMS-DL24 incorporate angular momentum loss mechanisms: the F06 prescription links mass flow to central star formation activity, while the HQ11 prescription bases inflow on gravitational torques from disk instabilities. Both feed a low-angular-momentum reservoir that subsequently accretes on viscous timescales. The GAEA-PMS-DL24 model further enhances black hole seeds by a factor of 50 relative to earlier implementations. Simba implements a dual-mode approach distinguishing between torque-limited accretion for cold gas and Bondi accretion for hot gas. Hydrodynamical simulations generally employ modified Bondi-Hoyle formulations: EAGLE includes angular momentum limitations through a viscous factor, IllustrisTNG incorporates magnetic pressure corrections to the sound speed.

        \textit{AGN Feedback Implementation.} AGN feedback modes differ fundamentally between preventative and ejective approaches. Radio mode feedback, ubiquitous across SAMs and present in EAGLE, operates by heating the hot gas halo to suppress cooling flows, with heating rates typically proportional to black hole accretion luminosity. Quasar-driven winds, implemented in GAEA-MS-F20F06/GAEA-MS-F20HQ11/GAEA-PMS-DL24, represent ejective feedback that directly removes cold gas from galaxies through kinetic outflows, with mass outflow rates proportional to black hole accretion rates (efficiency parameters $\epsilon_{qw} \sim 3$--5). IllustrisTNG implements a dual-mode framework where high-accretion states produce thermal (quasar-mode) feedback while low-accretion states trigger kinetic jet feedback-the latter being crucial for producing the observed red sequence and bimodality in galaxy colors. Simba similarly employs kinetic bipolar jets at low Eddington ratios ($f_{\rm Edd} < 0.02$) with velocities reaching $7000-8000$ km/s, supplemented by X-ray heating for gas-poor systems. The differences among different studies reflect the uncertainties of our current understanding of SMBH growth and feedback channels \citep{Shangguan2020,Yesuf2026}. 

        \subsection{Calibration Strategies and Observational Constraints}

        The different models employ diverse calibration philosophies that reflect different priorities in matching observational data. These strategies critically influence the parameter space explored and the properties (e.g. GSMF at z = 0-3) where models would be expected to perform well.

        \textit{Single-Epoch Calibration.} Several models calibrate primarily to $z\approx 0$ observables, relying on the intrinsic model physics to reproduce evolution at higher redshifts. LGal-ELUCID-Luo16 targets the local GSMF from \citet{Li2009} and \citet{Baldry2008}, with additional constraints from local HI and H$_2$ mass functions. JT1G-based GAEA variants (e.g., GAEA-JT1G-X20 and GAEA-JT1G-DL24) similarly focus on the $z=0$ GSMF from \citet{Li2009}, using HI/H$_2$ scaling relations as secondary constraints. Among hydrodynamical simulations, EAGLE calibrates to $z=0.1$ using the GSMF from \citet{Li2009} and \citet{Baldry2012}, alongside galaxy size-mass and black hole-galaxy mass relations. Simba uses \citet{Baldry2012} and \citet{Bernardi2017}.

        \textit{Multi-Epoch Calibration.} Several SAMs pursue more ambitious multi-redshift calibration strategies.  The GAEA-MS-H16FIRE/GAEA-MS-H16PREH/GAEA-MS-H16zDEP variants calibrate to GSMF evolution up to $z=3$, using observational compilations from \citet{Bell2003}, \citet{Drory2004}, \citet{Bundy2005}, \citet{Fontana2006}, and \citet{Ilbert2010,Ilbert2013}. GAEA-MS-F20F06 and GAEA-MS-F20HQ11 expand the constraint set by incorporating AGN luminosity functions from \citet{Ueda2014} and \citet{Fiore2012} alongside GSMF evolution, requiring simultaneous reproduction of both galaxy demographics and AGN activity. GAEA-PMS-DL24 further extends this approach to jointly constrain $z<3$ GSMF evolution, $z<4$ AGN luminosity functions, and local HI/H$_2$ mass functions, representing one of the most comprehensive calibration strategies among current SAMs. The TNG models mostly target $z=0$ constraints including the GSMF from \citet{Baldry2008,Baldry2012}, stellar mass-halo mass relations, and gas fractions in massive galaxy clusters, however, they also broadly constrain the model by using z = 0-3 GSMFs. The Simba simulations due to resolution limitations have to do a redshift dependent correction to match observations of high redshift SFRs and stellar masses.

        Multi-epoch calibration better constrains evolutionary pathways and feedback efficiencies, but introduces the risk that models may be over-tuned to specific datasets with their own systematic uncertainties. The substantial overlap in observational references used across models—particularly the widespread use of \citet{Li2009}, \citet{Baldry2012}, and \citet{Ilbert2010,Ilbert2013}—suggests that models have been tuned to similar observational SMFs. Thus, these models unavoidably will have inconsistencies with updated datasets (e.g. coming from JWST) or studies that consider other SPS/SFH assumptions that  are different than the ones they were constrained to reproduce.
        \end{multicols}

    \begin{table}[htbp]
        \centering
        \caption{Comparison of Galaxy Formation Models: Semi-analytic Models and Hydrodynamic Simulations}
        \label{tab:model_comparison}
        \vspace{2mm}

        \footnotesize
        \renewcommand{\arraystretch}{1.95}
        \setlength{\tabcolsep}{5pt}

        \begin{tabularx}{\textwidth}{>{\raggedright\arraybackslash}p{2.1cm} >{\raggedright\arraybackslash}p{1.0cm} c Z Z Z Z}
            \toprule
            \textit{Model}              & \textit{Box / Res.}     & \textit{Cali.\,(GSMF)} & \textit{SF Law}                                                                & \textit{Stellar Feedback}                                                    & \textit{BH Growth}                                                    & \textit{AGN Feedback}                                                           \\
            \midrule
            LGal-ELUCID-Luo16\citep{Luo2016}       & $L713m8.6$             & z=0                    & $\dot{M}_* \propto \Sigma_{\rm H_2}$; $H_2$ partition                      & SN reheating ($\eta \propto V_{\mathrm{vir}}^{-\beta}$); delayed reinc. + RPS     & Merger-driven quasar mode                                             & Radio-mode heating ($\dot{E} \propto \dot{M}_{\rm BH} c^2$)                     \\


            GAEA-MS-H16FIRE\citep{Hirschmann2016}        & $L685m9.1$             & z=0--3                 & $M_{\rm crit}$ threshold; bursty SF; non-instant recycling                     & Ejective energy-driven winds; metal-loaded                                   & Merger-driven cold gas accretion                                      & Radio-mode heating                                                              \\

            GAEA-MS-H16PREH\citep{Hirschmann2016}        & $L685m9.1$             & z=0--3                 & $M_{\rm crit}$ threshold; bursty SF; non-instant recycling                     & Preventative FB; halo pre-heating                                            & Same as above                                                         & Same as above                                                                   \\

            GAEA-MS-H16zDEP\citep{Hirschmann2016}        & $L685m9.1$             & z=0--3                 & $M_{\rm crit}$ threshold; bursty SF; non-instant recycling                     & $z$-dep ejection; $\tau_{\rm reinc} \propto t_{\rm dyn}$                     & Same as above                                                         & Same as above                                                                   \\

            GAEA-MS-F20F06\citep{Fontanot2020}         & $L685m9.1$             & z=0--3                 & Fixed $\alpha_{\rm SF}=0.1$; Chabrier IMF                                      & Ejective SN winds; fixed loading                                             & SFR-driven AM loss (flowJ)                                            & Quasar-mode winds ($\dot{M}_{\rm qw} \propto L_{\rm bol}$)                      \\

            GAEA-MS-F20HQ11\citep{Fontanot2020}           & $L685m9.1$             & z=0--3                 & Fixed $\alpha_{\rm SF}=0.1$; Chabrier IMF                                      & Ejective SN winds; fixed loading                                             & Torque-driven (HQ11)                                                  & Empirical outflows ($\dot{M}_{\rm qw} = 3.2 \dot{M}_{\rm BH}$)                  \\
            GAEA-PMS-DL24\citep{DeLucia2024}            & $L800m8.2$             & z=0--3                 & BR06 multi-ring $H_2$ partition; $H_2$-based SF                                & Ejective SN winds; mass-dep. reinc.; GRADHOT + cold-gas RPS                  & Reservoir-based (F06 torque)                                          & Dual: radio heating + quasar winds ($\dot{M}_{\rm qw} = 4.86 \dot{M}_{\rm BH}$) \\

            GAEA-JT1G-X20\citep{Xie2020,Han2025}          & $L1478m8.7$            & z=0                    & BR06 multi-ring $H_2$ partition; $H_2$-based SF                                & Ejective SN winds; mass-dep. reinc.; GRADHOT + cold-gas RPS                  & Boosted early seeding ($\times 50$)                                   & Radio-mode heating (as in DL24)                                                  \\
            GAEA-JT1G-DL24\citep{DeLucia2024}          & $L1478m8.7$            & z=0                    & Same SF law as GAEA-DL24/X20 (BR06 multi-ring $H_2$; $H_2$-based SF)            & Same feedback channel as GAEA-DL24/X20, including GRADHOT + cold-gas RPS      & Same BH-growth channel as GAEA-PMS-DL24                                    & Same AGN mode as GAEA-PMS-DL24                                                       \\

            \midrule\midrule

            TNG100-1\citep{Nelson2018}                & $L111m6.9/6.1$         & z=0                    & Effective EOS (SH03); density threshold $n_{\rm H} \approx 0.1\,{\rm cm}^{-3}$ & Kinetic winds; $v_w \propto \sigma, Z$; metal loading                        & Bondi accretion; MHD-modified ($c_{\rm s} \to c_{\rm s} + v_{\rm A}$) & Dual mode: Thermal (high-$\dot{m}$) + Kinetic (low-$\dot{m}$)                   \\

            TNG300-1\citep{Nelson2018}                & $L303m7.8/7.0$         & z=0                    & Same as above                                                                  & Same as above                                                                & Same as above                                                         & Same as above                                                                   \\

            TNG50-1\citep{Nelson2019}                 & $L52m5.7/4.9$          & z=0                    & High-$\rho$ SF: $t_* \propto \rho^{-1}$; $n_{\rm H} > 24.4\,{\rm cm}^{-3}$     & Resolved winds with natural collimation                                      & Bondi accretion; MHD-modified ($c_{\rm s} \to c_{\rm s} + v_{\rm A}$) & Ejective quenching via kinetic jets                                             \\

            Simba100\citep{Dave2019}                & $L147m8.1/7.4$         & z=0                    & $H_2$-based Schmidt law                                                        & 2-phase decoupled (hot+cold winds)                                           & Torque (cold) + Bondi (hot)                                           & Jets ($v_{\rm jet} \sim 7000$ km/s) + X-ray heating                             \\


            EAGLE100\citep{Schaye2015}                & $L100m7.0/6.3$         & z=0                    & Pressure-law SF ($\dot{M}_* \propto P^{1.5}$)                              & Stochastic thermal ($\Delta T = 10^{7.5}$ K)                                 & Mod. Bondi + AM limit                                                 & Single-mode thermal ($\Delta T = 10^{8.5}$ K)                                   \\
            \bottomrule
        \end{tabularx}
        \\[0.2em]
        \footnotesize
        \textit{Notes:} The \textit{Box / Res.} column gives the no-$h$ comoving box size and mass resolution; for hydrodynamical simulations, the two values are the dark-matter and stellar particle masses, separated by a slash. For SAMs, only the dark-matter particle mass is listed. Millennium-based models are abbreviated as MS, and their Millennium-II counterparts are not shown in the main table.
    \end{table}
\clearpage

\begin{multicols}{2}
        \section{Eddington Bias Correction and Analytical Models with Mass Cut}
        \label{sec:appendix_eb_masscut}

        In this supplementary section, we present the results of Eddington bias correction with a mass cut at $\log(M_*/M_\odot) = 11.75$. This mass threshold is applied to ensure the reliability of our stellar mass estimates and to minimize the impact of large photometric uncertainties at the extreme high-mass end.

\begin{figure}[H]
    \centering
    \includegraphics[width=0.8\textwidth]{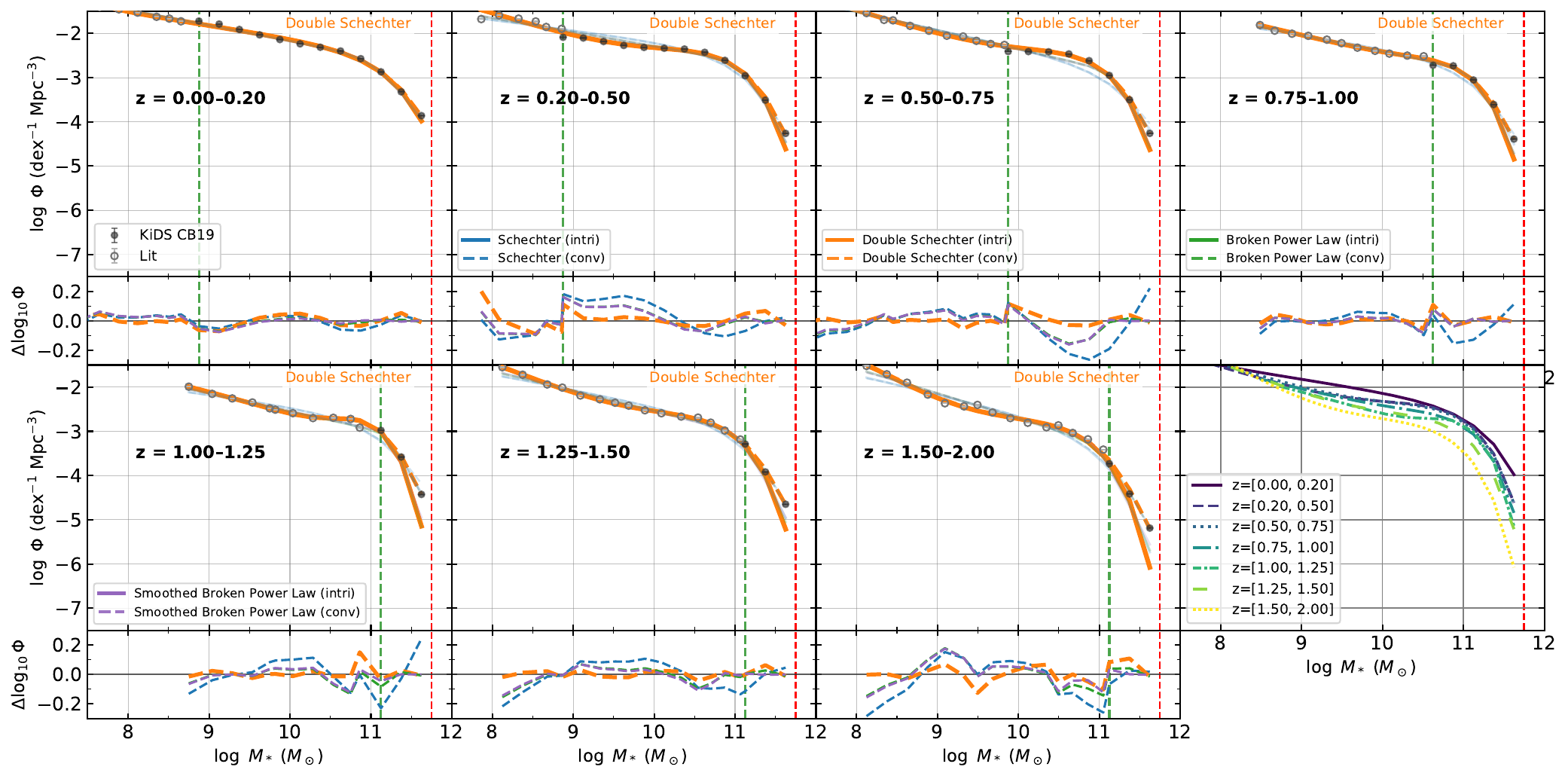}
    \caption{Stellar mass functions with Eddington bias correction using direct fitting with a mass cut at $\log(M_*/M_\odot) = 11.75$. Different analytical models are compared using the Gaussian-Lorentzian kernel choice following \citet{Weaver2023}.}
    \label{fig:smf_eb_cutmass}
\end{figure}

        Figure~\ref{fig:smf_eb_cutmass} shows the comparison of different analytical model fits (single Schechter, double Schechter, broken power law and smoothed broken power law) to the observed GSMF after applying the mass cut, using the Gaussian-Lorentzian kernel choice following \citet{Weaver2023} as in the main text. The correction can reduces the number densities for the extreme high-mass end by even 1 dex, consistent with expectations from Eddington bias effects.

\begin{figure}[H]
    \centering
    \includegraphics[width=0.8\textwidth]{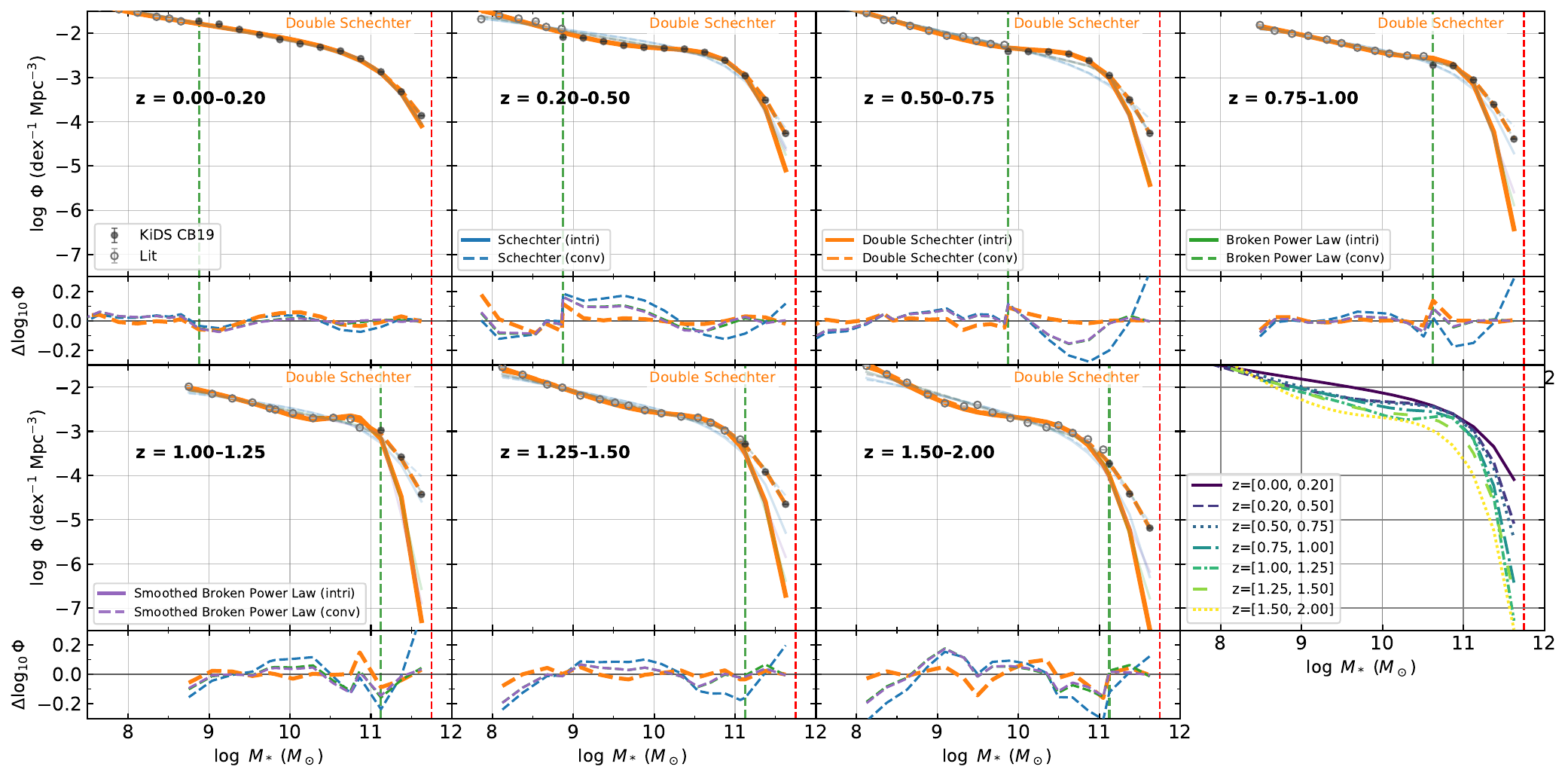}
    \caption{Stellar mass functions with Eddington bias correction using direct fitting with a mass cut at $\log(M_*/M_\odot) = 11.75$. Different analytical models are compared using the redshift-dependent Gaussian kernel choice following \citet{Chaikin2025}.}
    \label{fig:smf_eb_cutmass_colibre}
\end{figure}

        Figure~\ref{fig:smf_eb_cutmass_colibre} presents the same analysis using the redshift-dependent Gaussian kernel choice following \citet{Chaikin2025}, as in the main text. Unlike the no-mass-cut case discussed in the main text, after applying the mass cut the double Schechter function is preferred in all redshift bins; the corresponding best-fit parameters are listed in Table~\ref{tab:fit_parameters_cut}.

\end{multicols}

\clearpage

\begin{multicols}{2}
\section{GSMF Results table}
\label{sec:gsmf_table}
\end{multicols}

\begin{table}[htbp]
\centering
\caption{GSMF results for the CIGALE CB19 (CICB19) stellar masses. The first header row lists the redshift bins, while the second header row gives the corresponding correction factors. Here $\Phi_{\mathrm{out.fr}}$ denotes the correction for removing the outlier fraction by retaining sources with $|\delta z|<0.10$, as described in the main text, and $\Phi_{z,\mathrm{cor}}$ denotes the correction caused by GaZNet redshift scattering between redshift bins, as described in the main text Each subsequent row reports the finally corrected logarithmic GSMF value in a stellar-mass bin, where the mass is given as $\log(M_*/M_\odot)$ and the GSMF has units of $\mathrm{Mpc}^{-3}\,\mathrm{dex}^{-1}$.}
\setlength{\tabcolsep}{3pt}
\begin{tabular}{c|c|ccccccc}
\hline
 $\Phi$& z & $0.001-0.2$ & $0.2-0.5$ & $0.5-0.75$ & $0.75-1.0$ & $1.0-1.25$ & $1.25-1.5$ & $1.5-2.0$ \\
 \hline
 $\log M_*/M_\odot$& \diagbox[width=5.6em,height=2.6em]{$\Phi_{\mathrm{out.fr}}$}{$\Phi_{z,\mathrm{cor}}$} & $0.1629$ & $0.0055$ & $-0.0212$ & $-0.0983$ & $-0.0760$ & $0.0243$ & $0.0540$ \\
\hline
$8.125$ & $0.0000$ & $-1.783^{+0.005}_{-0.005}$ & - & - & - & - & - & - \\
$8.375$ & $0.0000$ & $-1.748^{+0.002}_{-0.002}$ & - & - & - & - & - & - \\
$8.625$ & $0.0000$ & $-1.758^{+0.001}_{-0.001}$ &  -  & - & - & - & - & - \\
$8.875$ & $-0.0018$ & $-1.723^{+0.001}_{-0.001}$ & $-2.090^{+0.002}_{-0.002}$ & - & - & - & - & - \\
$9.125$ & $-0.0019$ & $-1.790^{+0.001}_{-0.001}$ & $-2.106^{+0.001}_{-0.001}$ & - & - & - & - & - \\
$9.375$ & $-0.0031$ & $-1.914^{+0.002}_{-0.002}$ & $-2.187^{+0.001}_{-0.001}$ & - & - & - & - & - \\
$9.625$ & $-0.0032$ & $-2.037^{+0.002}_{-0.002}$ & $-2.276^{+0.001}_{-0.001}$ & - & - & - & - & - \\
$9.875$ & $-0.0028$ & $-2.141^{+0.002}_{-0.002}$ & $-2.320^{+0.001}_{-0.001}$ & $-2.412^{+0.002}_{-0.002}$ & - & - & - & - \\
$10.125$ & $-0.0026$ & $-2.239^{+0.002}_{-0.002}$ & $-2.338^{+0.001}_{-0.001}$ & $-2.414^{+0.001}_{-0.001}$ & - & - & - & - \\
$10.375$ & $-0.0028$ & $-2.316^{+0.002}_{-0.002}$ & $-2.356^{+0.001}_{-0.001}$ & $-2.419^{+0.001}_{-0.001}$ & - & - & - & - \\
$10.625$ & $-0.0032$ & $-2.400^{+0.003}_{-0.003}$ & $-2.431^{+0.001}_{-0.001}$ & $-2.468^{+0.001}_{-0.001}$ & $-2.724^{+0.003}_{-0.003}$ & - & - & - \\
$10.875$ & $-0.0034$ & $-2.575^{+0.003}_{-0.003}$ & $-2.613^{+0.001}_{-0.001}$ & $-2.620^{+0.001}_{-0.001}$ & $-2.736^{+0.001}_{-0.001}$ & - & - & - \\
$11.125$ & $-0.0041$ & $-2.874^{+0.005}_{-0.005}$ & $-2.961^{+0.001}_{-0.001}$ & $-2.955^{+0.001}_{-0.001}$ & $-3.059^{+0.001}_{-0.001}$ & $-2.983^{+0.001}_{-0.001}$ & $-3.292^{+0.002}_{-0.002}$ & $-3.735^{+0.002}_{-0.002}$ \\
$11.375$ & $-0.0061$ & $-3.327^{+0.008}_{-0.008}$ & $-3.515^{+0.002}_{-0.002}$ & $-3.507^{+0.002}_{-0.002}$ & $-3.614^{+0.002}_{-0.002}$ & $-3.587^{+0.003}_{-0.003}$ & $-3.924^{+0.005}_{-0.005}$ & $-4.422^{+0.005}_{-0.005}$ \\
$11.625$ & $-0.0117$ & $-3.872^{+0.015}_{-0.015}$ & $-4.274^{+0.006}_{-0.006}$ & $-4.272^{+0.004}_{-0.004}$ & $-4.401^{+0.005}_{-0.005}$ & $-4.438^{+0.006}_{-0.006}$ & $-4.662^{+0.011}_{-0.011}$ & $-5.199^{+0.012}_{-0.012}$ \\
$11.875$ & $-0.0450$ & $-4.490^{+0.029}_{-0.030}$ & $-5.299^{+0.018}_{-0.018}$ & $-5.341^{+0.014}_{-0.014}$ & $-5.463^{+0.014}_{-0.014}$ & $-5.315^{+0.018}_{-0.018}$ & $-5.252^{+0.024}_{-0.024}$ & $-5.959^{+0.032}_{-0.032}$ \\
$12.125$ & $-0.1631$ & $-5.068^{+0.051}_{-0.051}$ & $-6.377^{+0.057}_{-0.058}$ & $-6.470^{+0.045}_{-0.045}$ & $-6.457^{+0.038}_{-0.038}$ & $-5.996^{+0.037}_{-0.037}$ & $-5.965^{+0.053}_{-0.054}$ & $-6.779^{+0.063}_{-0.064}$ \\
$12.375$ & $-0.3174$ & $-5.804^{+0.104}_{-0.106}$ & $-7.192^{+0.129}_{-0.134}$ & $-7.245^{+0.094}_{-0.096}$ & $-7.076^{+0.060}_{-0.061}$ & $-6.505^{+0.067}_{-0.068}$ & $-6.681^{+0.100}_{-0.103}$ & $-7.417^{+0.123}_{-0.127}$ \\
\hline
\end{tabular}
\label{tab:gamf_table}
\end{table}

\begin{multicols}{2}
    
\section{Best-Fit Parameters for Analytical Models}
\label{sec:appendix_fit_parameters}
In this supplementary section, we provide the best-fit parameters for all analytical models across different redshift bins. Table~\ref{tab:fit_parameters_nocut} presents the results without mass cut, while Table~\ref{tab:fit_parameters_cut} shows the results with mass cut at $\log(M_*/M_\odot) = 11.75$. In each cell, the Chaikin2025-based results are shown in the top line, and the Weaver2023-based results are shown in the bottom line. All values include their $1\sigma$ uncertainties as superscripts.

For the no-mass-cut case discussed in the main text, Table~\ref{tab:fit_parameters_nocut} shows that the DS fits generally provide flexible descriptions of the GSMF, but their uncertainties are large in the lowest redshift bin ($z=0$--0.2), especially for the second power-law/Schechter component. This likely reflects the additional leverage introduced by the newly included high-mass-end data, which makes the decomposition between the two components less uniquely constrained. For the SBPL fits, the smoothness parameter $s$ becomes close to zero at $z>0.5$, indicating that the model approaches the BPL limit. Nevertheless, the additional freedom introduced by $s$ still allows SBPL to fit the data slightly better than BPL in several redshift bins, as reflected by its smaller MAD values.

For the mass-cut case discussed in Section~\ref{sec:appendix_eb_masscut}, Table~\ref{tab:fit_parameters_cut} shows that the DS form provides the preferred description after imposing $\log(M_*/M_\odot)<11.75$. The parameter uncertainties are generally reduced relative to the no-mass-cut case, although sizeable uncertainties remain in some redshift bins. For the BPL fits, the high-mass slope $\beta$ is often weakly constrained and has large uncertainties. Similarly, in the SBPL fits, $\beta$ remains uncertain and $s$ is again close to zero, suggesting a shape close to BPL; however, the extra smoothness freedom still improves the fit to the data points, leading to smaller MAD values in several bins.

We note that for the DS fit the large error bars just reflect that the functional form is not an optimal choice to describe the high mass data unlike the SBPL, while they have the same number of parameters.

\end{multicols}
\newcommand{\stackval}[2]{%
    \begin{tabular}[t]{@{}c@{}}
        $#1$ \\
        $#2$
    \end{tabular}%
}


\begin{table}[htbp]
    \centering
    \caption{Best-fit parameters for analytical models without mass cut. Each cell shows COLIBRE results (top) and Weaver2023 results (bottom) with $1\sigma$ uncertainties. Model abbreviations: SS (Schechter), DS (Double Schechter), BPL (Broken Power Law), SBPL (Smoothed Broken Power Law).}
    \label{tab:fit_parameters_nocut}
    
    \footnotesize
    \renewcommand{\arraystretch}{1.4}
    \setlength{\tabcolsep}{3pt}
    
    \begin{tabular}{llccccccc}
        \toprule
                Model & Parameter & $[0.0, 0.2)$ & $ [0.2, 0.5)$ & $[0.5, 0.75)$ & $[0.75, 1.0)$ & $[1.0, 1.25)$ & $[1.25, 1.5)$ &$[1.5, 2.0)$\\
        \midrule

        \multirow{4}{*}{SS}
        & $\log_{10}\phi^*$  & \stackval{-3.044^{+0.192}_{-0.192}}{-3.055^{+0.188}_{-0.188}}  & \stackval{-2.960^{+0.268}_{-0.268}}{-2.969^{+0.304}_{-0.304}}  & \stackval{-3.104^{+0.235}_{-0.235}}{-3.139^{+0.231}_{-0.231}}  & \stackval{-2.991^{+0.209}_{-0.209}}{-3.104^{+0.180}_{-0.180}}  & \stackval{-3.131^{+0.385}_{-0.385}}{-3.349^{+0.238}_{-0.238}}  & \stackval{-3.326^{+0.379}_{-0.379}}{-3.452^{+0.234}_{-0.234}}  & \stackval{-3.593^{+0.271}_{-0.271}}{-3.774^{+0.227}_{-0.227}}  \\
        & $\log_{10}M^*$  & \stackval{11.392^{+0.070}_{-0.070}}{11.422^{+0.067}_{-0.067}}  & \stackval{11.072^{+0.069}_{-0.069}}{11.121^{+0.079}_{-0.079}}  & \stackval{11.059^{+0.106}_{-0.106}}{11.132^{+0.071}_{-0.071}}  & \stackval{10.975^{+0.082}_{-0.082}}{11.108^{+0.058}_{-0.058}}  & \stackval{11.069^{+0.140}_{-0.140}}{11.252^{+0.083}_{-0.083}}  & \stackval{11.129^{+0.120}_{-0.120}}{11.281^{+0.078}_{-0.078}}  & \stackval{11.021^{+0.088}_{-0.088}}{11.187^{+0.071}_{-0.071}}  \\
        & $\alpha$  & \stackval{-1.362^{+0.068}_{-0.068}}{-1.362^{+0.066}_{-0.066}}  & \stackval{-1.310^{+0.121}_{-0.121}}{-1.306^{+0.137}_{-0.137}}  & \stackval{-1.399^{+0.103}_{-0.103}}{-1.403^{+0.104}_{-0.104}}  & \stackval{-1.300^{+0.108}_{-0.108}}{-1.335^{+0.093}_{-0.093}}  & \stackval{-1.291^{+0.208}_{-0.208}}{-1.376^{+0.117}_{-0.117}}  & \stackval{-1.416^{+0.154}_{-0.154}}{-1.445^{+0.101}_{-0.101}}  & \stackval{-1.544^{+0.136}_{-0.136}}{-1.589^{+0.105}_{-0.105}}  \\
        & MAD  & \stackval{0.116}{0.119}  & \stackval{0.147}{0.137}  & \stackval{0.126}{0.124}  & \stackval{0.116}{0.115}  & \stackval{0.154}{0.158}  & \stackval{0.126}{0.137}  & \stackval{0.163}{0.175}  \\
        \hline

        \multirow{6}{*}{DS}
        & $\log_{10}\phi_1^*$  & \stackval{-3.500^{+42.751}_{-42.751}}{-3.500^{+38.569}_{-38.569}}  & \stackval{-2.805^{+0.296}_{-0.296}}{-2.851^{+0.267}_{-0.267}}  & \stackval{-2.812^{+0.388}_{-0.388}}{-2.824^{+0.361}_{-0.361}}  & \stackval{-2.902^{+0.242}_{-0.242}}{-3.299^{+31.597}_{-31.597}}  & \stackval{-3.024^{+0.352}_{-0.352}}{-3.188^{+0.560}_{-0.560}}  & \stackval{-3.127^{+0.241}_{-0.241}}{-3.327^{+0.404}_{-0.404}}  & \stackval{-3.259^{+0.601}_{-0.601}}{-3.370^{+0.242}_{-0.242}}  \\
        & $\log_{10}\phi_2^*$  & \stackval{-3.166^{+21.561}_{-21.561}}{-3.180^{+20.005}_{-20.005}}  & \stackval{-4.241^{+6.775}_{-6.775}}{-4.227^{+6.942}_{-6.942}}  & \stackval{-4.125^{+2.179}_{-2.179}}{-3.954^{+1.764}_{-1.764}}  & \stackval{-5.000^{+12.797}_{-12.797}}{-3.468^{+48.147}_{-48.147}}  & \stackval{-5.000^{+8.502}_{-8.502}}{-5.000^{+12.150}_{-12.150}}  & \stackval{-5.000^{+5.110}_{-5.110}}{-4.267^{+5.477}_{-5.477}}  & \stackval{-5.000^{+5.169}_{-5.169}}{-4.931^{+2.828}_{-2.828}}  \\
        & $\log_{10}M^*$  & \stackval{11.376^{+0.652}_{-0.652}}{11.406^{+0.570}_{-0.570}}  & \stackval{11.020^{+0.136}_{-0.136}}{11.075^{+0.130}_{-0.130}}  & \stackval{10.960^{+0.139}_{-0.139}}{11.025^{+0.139}_{-0.139}}  & \stackval{10.940^{+0.132}_{-0.132}}{11.095^{+0.417}_{-0.417}}  & \stackval{11.015^{+0.188}_{-0.188}}{11.194^{+0.267}_{-0.267}}  & \stackval{11.049^{+0.138}_{-0.138}}{11.214^{+0.198}_{-0.198}}  & \stackval{10.901^{+0.263}_{-0.263}}{11.059^{+0.123}_{-0.123}}  \\
        & $\alpha_1$  & \stackval{-1.200^{+17.602}_{-17.602}}{-1.200^{+15.822}_{-15.822}}  & \stackval{-1.100^{+0.849}_{-0.849}}{-1.122^{+0.834}_{-0.834}}  & \stackval{-1.099^{+0.573}_{-0.573}}{-1.060^{+0.602}_{-0.602}}  & \stackval{-1.180^{+0.629}_{-0.629}}{-1.200^{+12.985}_{-12.985}}  & \stackval{-1.097^{+0.846}_{-0.846}}{-1.200^{+1.151}_{-1.151}}  & \stackval{-1.200^{+0.506}_{-0.506}}{-1.200^{+1.286}_{-1.286}}  & \stackval{-1.175^{+1.045}_{-1.045}}{-1.200^{+0.515}_{-0.515}}  \\
        & $\alpha_2$  & \stackval{-1.385^{+2.963}_{-2.963}}{-1.386^{+2.766}_{-2.766}}  & \stackval{-1.729^{+1.974}_{-1.974}}{-1.708^{+1.963}_{-1.963}}  & \stackval{-1.732^{+0.564}_{-0.564}}{-1.670^{+0.447}_{-0.447}}  & \stackval{-2.045^{+4.824}_{-4.824}}{-1.428^{+9.076}_{-9.076}}  & \stackval{-2.120^{+3.425}_{-3.425}}{-1.989^{+4.190}_{-4.190}}  & \stackval{-2.029^{+1.677}_{-1.677}}{-1.728^{+1.525}_{-1.525}}  & \stackval{-2.116^{+1.755}_{-1.755}}{-2.028^{+0.912}_{-0.912}}  \\
        & MAD  & \stackval{0.117}{0.121}  & \stackval{0.143}{0.132}  & \stackval{0.092}{0.093}  & \stackval{0.111}{0.116}  & \stackval{0.138}{0.155}  & \stackval{0.105}{0.131}  & \stackval{0.145}{0.167}  \\
        \hline

        \multirow{5}{*}{BPL}
        & $\log_{10}\phi^*$  & \stackval{-2.521^{+0.201}_{-0.201}}{-2.525^{+0.199}_{-0.199}}  & \stackval{-2.633^{+0.271}_{-0.271}}{-2.657^{+0.235}_{-0.235}}  & \stackval{-2.831^{+0.265}_{-0.265}}{-2.863^{+0.251}_{-0.251}}  & \stackval{-2.769^{+0.235}_{-0.235}}{-2.823^{+0.237}_{-0.237}}  & \stackval{-2.872^{+0.582}_{-0.582}}{-2.952^{+0.638}_{-0.638}}  & \stackval{-2.931^{+0.418}_{-0.418}}{-2.957^{+0.381}_{-0.381}}  & \stackval{-3.196^{+0.500}_{-0.500}}{-3.232^{+0.435}_{-0.435}}  \\
        & $\log_{10}M^*$  & \stackval{11.051^{+0.177}_{-0.177}}{11.063^{+0.175}_{-0.175}}  & \stackval{11.102^{+0.139}_{-0.139}}{11.140^{+0.136}_{-0.136}}  & \stackval{11.171^{+0.119}_{-0.119}}{11.229^{+0.130}_{-0.130}}  & \stackval{11.070^{+0.126}_{-0.126}}{11.157^{+0.130}_{-0.130}}  & \stackval{11.054^{+0.605}_{-0.605}}{11.148^{+0.709}_{-0.709}}  & \stackval{10.983^{+0.518}_{-0.518}}{11.044^{+0.431}_{-0.431}}  & \stackval{10.911^{+0.437}_{-0.437}}{10.980^{+0.291}_{-0.291}}  \\
        & $\alpha$  & \stackval{-0.344^{+0.068}_{-0.068}}{-0.345^{+0.068}_{-0.068}}  & \stackval{-0.315^{+0.118}_{-0.118}}{-0.321^{+0.109}_{-0.109}}  & \stackval{-0.414^{+0.105}_{-0.105}}{-0.418^{+0.096}_{-0.096}}  & \stackval{-0.349^{+0.111}_{-0.111}}{-0.362^{+0.109}_{-0.109}}  & \stackval{-0.345^{+0.233}_{-0.233}}{-0.373^{+0.220}_{-0.220}}  & \stackval{-0.425^{+0.131}_{-0.131}}{-0.427^{+0.126}_{-0.126}}  & \stackval{-0.545^{+0.228}_{-0.228}}{-0.547^{+0.218}_{-0.218}}  \\
        & $\beta$  & \stackval{-2.362^{+0.297}_{-0.297}}{-2.329^{+0.291}_{-0.291}}  & \stackval{-3.877^{+0.611}_{-0.611}}{-3.799^{+0.775}_{-0.775}}  & \stackval{-4.327^{+0.691}_{-0.691}}{-4.216^{+0.915}_{-0.915}}  & \stackval{-4.032^{+0.669}_{-0.669}}{-3.887^{+0.815}_{-0.815}}  & \stackval{-3.286^{+1.931}_{-1.931}}{-3.153^{+1.968}_{-1.968}}  & \stackval{-2.924^{+1.381}_{-1.381}}{-2.804^{+0.963}_{-0.963}}  & \stackval{-3.314^{+1.243}_{-1.243}}{-3.187^{+0.640}_{-0.640}}  \\
        & MAD  & \stackval{0.050}{0.049}  & \stackval{0.066}{0.068}  & \stackval{0.069}{0.067}  & \stackval{0.058}{0.053}  & \stackval{0.096}{0.089}  & \stackval{0.074}{0.067}  & \stackval{0.104}{0.095}  \\
        \hline

        \multirow{6}{*}{SBPL}
        & $\log_{10}\phi^*$  & \stackval{-2.529^{+0.138}_{-0.138}}{-2.535^{+0.140}_{-0.140}}  & \stackval{-2.640^{+0.222}_{-0.222}}{-2.660^{+0.235}_{-0.235}}  & \stackval{-2.824^{+0.230}_{-0.230}}{-2.870^{+0.233}_{-0.233}}  & \stackval{-2.773^{+0.196}_{-0.196}}{-2.822^{+0.255}_{-0.255}}  & \stackval{-2.880^{+0.232}_{-0.232}}{-3.017^{+0.263}_{-0.263}}  & \stackval{-2.940^{+0.246}_{-0.246}}{-2.979^{+0.210}_{-0.210}}  & \stackval{-3.207^{+0.251}_{-0.251}}{-3.181^{+0.247}_{-0.247}}  \\
        & $\log_{10}M^*$  & \stackval{10.982^{+0.142}_{-0.142}}{10.998^{+0.147}_{-0.147}}  & \stackval{11.062^{+0.251}_{-0.251}}{11.129^{+0.255}_{-0.255}}  & \stackval{11.092^{+0.107}_{-0.107}}{11.181^{+0.209}_{-0.209}}  & \stackval{11.023^{+0.091}_{-0.091}}{11.104^{+0.419}_{-0.419}}  & \stackval{10.942^{+0.190}_{-0.190}}{11.125^{+0.177}_{-0.177}}  & \stackval{10.900^{+0.305}_{-0.305}}{10.954^{+0.142}_{-0.142}}  & \stackval{10.845^{+0.123}_{-0.123}}{10.879^{+0.131}_{-0.131}}  \\
        & $\alpha$  & \stackval{-0.357^{+0.056}_{-0.056}}{-0.357^{+0.056}_{-0.056}}  & \stackval{-0.325^{+0.121}_{-0.121}}{-0.324^{+0.126}_{-0.126}}  & \stackval{-0.425^{+0.096}_{-0.096}}{-0.430^{+0.095}_{-0.095}}  & \stackval{-0.360^{+0.100}_{-0.100}}{-0.372^{+0.123}_{-0.123}}  & \stackval{-0.373^{+0.130}_{-0.130}}{-0.413^{+0.128}_{-0.128}}  & \stackval{-0.446^{+0.092}_{-0.092}}{-0.458^{+0.097}_{-0.097}}  & \stackval{-0.574^{+0.141}_{-0.141}}{-0.548^{+0.165}_{-0.165}}  \\
        & $\beta$  & \stackval{-2.215^{+0.309}_{-0.309}}{-2.194^{+0.297}_{-0.297}}  & \stackval{-3.635^{+1.459}_{-1.459}}{-3.731^{+1.557}_{-1.557}}  & \stackval{-3.805^{+0.708}_{-0.708}}{-3.825^{+1.815}_{-1.815}}  & \stackval{-3.758^{+0.608}_{-0.608}}{-3.504^{+3.221}_{-3.221}}  & \stackval{-2.854^{+0.588}_{-0.588}}{-2.999^{+0.478}_{-0.478}}  & \stackval{-2.673^{+0.627}_{-0.627}}{-2.540^{+0.277}_{-0.277}}  & \stackval{-3.073^{+0.301}_{-0.301}}{-2.927^{+0.274}_{-0.274}}  \\
        & $s$  & \stackval{0.257^{+0.429}_{-0.429}}{0.285^{+0.415}_{-0.415}}  & \stackval{0.178^{+0.768}_{-0.768}}{0.259^{+0.648}_{-0.648}}  & \stackval{0.004}{0.151^{+0.449}_{-0.449}}  & \stackval{0.011}{0.177^{+0.820}_{-0.820}}  & \stackval{0.009}{0.000}  & \stackval{0.000}{0.000}  & \stackval{0.001}{0.000}  \\
        & MAD  & \stackval{0.042}{0.042}  & \stackval{0.063}{0.068}  & \stackval{0.061}{0.063}  & \stackval{0.046}{0.052}  & \stackval{0.080}{0.068}  & \stackval{0.051}{0.045}  & \stackval{0.079}{0.069}  \\
        \hline
\end{tabular}
\end{table}

\begin{table}[htbp]
    \centering
    \caption{Best-fit parameters for analytical models with mass cut at $\log(M_*/M_\odot) = 11.75$. Each cell shows COLIBRE results (top) and Weaver2023 results (bottom) with $1\sigma$ uncertainties. Model abbreviations: SS (Schechter), DS (Double Schechter), BPL (Broken Power Law), SBPL (Smoothed Broken Power Law).}
    \label{tab:fit_parameters_cut}
    
    \footnotesize
    \renewcommand{\arraystretch}{1.4}
    \setlength{\tabcolsep}{3pt}
    
    \begin{tabular}{llccccccc}
        \toprule
                Model & Parameter & $[0.0, 0.2)$ & $ [0.2, 0.5)$ & $[0.5, 0.75)$ & $[0.75, 1.0)$ & $[1.0, 1.25)$ & $[1.25, 1.5)$ &$[1.5, 2.0)$\\
        \midrule

        \multirow{4}{*}{SS}
        & $\log_{10}\phi^*$  & \stackval{-2.839^{+0.200}_{-0.200}}{-2.861^{+0.195}_{-0.195}}  & \stackval{-2.764^{+0.361}_{-0.361}}{-2.773^{+0.272}_{-0.272}}  & \stackval{-3.161^{+0.313}_{-0.313}}{-3.160^{+0.306}_{-0.306}}  & \stackval{-2.983^{+0.303}_{-0.303}}{-3.000^{+0.293}_{-0.293}}  & \stackval{-3.049^{+0.407}_{-0.407}}{-3.089^{+0.368}_{-0.368}}  & \stackval{-3.114^{+0.361}_{-0.361}}{-3.135^{+0.442}_{-0.442}}  & \stackval{-3.302^{+0.480}_{-0.480}}{-3.360^{+0.451}_{-0.451}}  \\
        & $\log_{10}M^*$  & \stackval{11.113^{+0.104}_{-0.104}}{11.148^{+0.098}_{-0.098}}  & \stackval{10.957^{+0.194}_{-0.194}}{10.988^{+0.122}_{-0.122}}  & \stackval{11.162^{+0.228}_{-0.228}}{11.185^{+0.223}_{-0.223}}  & \stackval{10.984^{+0.209}_{-0.209}}{11.034^{+0.213}_{-0.213}}  & \stackval{11.021^{+0.285}_{-0.285}}{11.080^{+0.315}_{-0.315}}  & \stackval{10.905^{+0.214}_{-0.214}}{10.972^{+0.209}_{-0.209}}  & \stackval{10.787^{+0.241}_{-0.241}}{10.881^{+0.173}_{-0.173}}  \\
        & $\alpha$  & \stackval{-1.322^{+0.077}_{-0.077}}{-1.327^{+0.076}_{-0.076}}  & \stackval{-1.240^{+0.132}_{-0.132}}{-1.239^{+0.104}_{-0.104}}  & \stackval{-1.405^{+0.110}_{-0.110}}{-1.402^{+0.110}_{-0.110}}  & \stackval{-1.294^{+0.136}_{-0.136}}{-1.294^{+0.129}_{-0.129}}  & \stackval{-1.250^{+0.221}_{-0.221}}{-1.264^{+0.189}_{-0.189}}  & \stackval{-1.359^{+0.155}_{-0.155}}{-1.358^{+0.186}_{-0.186}}  & \stackval{-1.430^{+0.314}_{-0.314}}{-1.440^{+0.290}_{-0.290}}  \\
        & MAD  & \stackval{0.056}{0.055}  & \stackval{0.091}{0.083}  & \stackval{0.095}{0.088}  & \stackval{0.068}{0.054}  & \stackval{0.103}{0.088}  & \stackval{0.101}{0.083}  & \stackval{0.129}{0.110}  \\
        \hline

        \multirow{6}{*}{DS}
        & $\log_{10}\phi_1^*$  & \stackval{-2.711^{+0.224}_{-0.224}}{-2.745^{+0.229}_{-0.229}}  & \stackval{-2.466^{+0.188}_{-0.188}}{-2.550^{+0.231}_{-0.231}}  & \stackval{-2.463^{+0.155}_{-0.155}}{-2.602^{+0.226}_{-0.226}}  & \stackval{-2.563^{+0.272}_{-0.272}}{-2.762^{+0.362}_{-0.362}}  & \stackval{-2.680^{+0.326}_{-0.326}}{-2.746^{+0.242}_{-0.242}}  & \stackval{-2.687^{+0.180}_{-0.180}}{-2.869^{+0.191}_{-0.191}}  & \stackval{-2.847^{+0.200}_{-0.200}}{-3.022^{+0.264}_{-0.264}}  \\
        & $\log_{10}\phi_2^*$  & \stackval{-4.313^{+5.578}_{-5.578}}{-4.171^{+5.774}_{-5.774}}  & \stackval{-3.362^{+0.898}_{-0.898}}{-3.499^{+1.322}_{-1.322}}  & \stackval{-3.240^{+0.438}_{-0.438}}{-3.457^{+0.605}_{-0.605}}  & \stackval{-3.172^{+1.068}_{-1.068}}{-3.189^{+0.910}_{-0.910}}  & \stackval{-3.456^{+1.856}_{-1.856}}{-3.409^{+0.815}_{-0.815}}  & \stackval{-3.409^{+0.691}_{-0.691}}{-3.872^{+1.407}_{-1.407}}  & \stackval{-4.384^{+2.012}_{-2.012}}{-4.454^{+1.987}_{-1.987}}  \\
        & $\log_{10}M^*$  & \stackval{11.044^{+0.175}_{-0.175}}{11.082^{+0.200}_{-0.200}}  & \stackval{10.723^{+0.180}_{-0.180}}{10.830^{+0.203}_{-0.203}}  & \stackval{10.658^{+0.168}_{-0.168}}{10.838^{+0.154}_{-0.154}}  & \stackval{10.592^{+0.383}_{-0.383}}{10.783^{+0.234}_{-0.234}}  & \stackval{10.618^{+0.822}_{-0.822}}{10.733^{+0.296}_{-0.296}}  & \stackval{10.538^{+0.206}_{-0.206}}{10.804^{+0.163}_{-0.163}}  & \stackval{10.540^{+0.174}_{-0.174}}{10.720^{+0.177}_{-0.177}}  \\
        & $\alpha_1$  & \stackval{-1.200^{+0.458}_{-0.458}}{-1.200^{+0.558}_{-0.558}}  & \stackval{-0.472^{+0.853}_{-0.853}}{-0.659^{+0.893}_{-0.893}}  & \stackval{-0.269^{+0.883}_{-0.883}}{-0.674^{+0.636}_{-0.636}}  & \stackval{-0.100^{+1.968}_{-1.968}}{-0.343^{+1.553}_{-1.553}}  & \stackval{-0.100^{+3.145}_{-3.145}}{-0.100^{+1.603}_{-1.603}}  & \stackval{-0.157^{+1.194}_{-1.194}}{-0.802^{+0.797}_{-0.797}}  & \stackval{-0.742^{+0.881}_{-0.881}}{-0.918^{+0.810}_{-0.810}}  \\
        & $\alpha_2$  & \stackval{-1.695^{+1.229}_{-1.229}}{-1.654^{+1.235}_{-1.235}}  & \stackval{-1.534^{+0.337}_{-0.337}}{-1.561^{+0.456}_{-0.456}}  & \stackval{-1.527^{+0.151}_{-0.151}}{-1.567^{+0.182}_{-0.182}}  & \stackval{-1.487^{+0.510}_{-0.510}}{-1.439^{+0.414}_{-0.414}}  & \stackval{-1.614^{+0.811}_{-0.811}}{-1.540^{+0.378}_{-0.378}}  & \stackval{-1.610^{+0.305}_{-0.305}}{-1.728^{+0.511}_{-0.511}}  & \stackval{-2.055^{+0.841}_{-0.841}}{-1.997^{+0.716}_{-0.716}}  \\
        & MAD  & \stackval{0.041}{0.043}  & \stackval{0.038}{0.047}  & \stackval{0.024}{0.024}  & \stackval{0.031}{0.026}  & \stackval{0.050}{0.028}  & \stackval{0.023}{0.021}  & \stackval{0.049}{0.047}  \\
        \hline

        \multirow{5}{*}{BPL}
        & $\log_{10}\phi^*$  & \stackval{-2.567^{+0.177}_{-0.177}}{-2.566^{+0.181}_{-0.181}}  & \stackval{-2.646^{+0.282}_{-0.282}}{-2.651^{+0.272}_{-0.272}}  & \stackval{-2.838^{+0.288}_{-0.288}}{-2.885^{+0.262}_{-0.262}}  & \stackval{-2.786^{+0.247}_{-0.247}}{-2.845^{+0.275}_{-0.275}}  & \stackval{-2.956^{+0.637}_{-0.637}}{-3.023^{+0.466}_{-0.466}}  & \stackval{-2.978^{+0.256}_{-0.256}}{-3.053^{+0.263}_{-0.263}}  & \stackval{-3.250^{+0.440}_{-0.440}}{-3.311^{+0.329}_{-0.329}}  \\
        & $\log_{10}M^*$  & \stackval{11.116^{+0.161}_{-0.161}}{11.120^{+0.167}_{-0.167}}  & \stackval{11.113^{+0.165}_{-0.165}}{11.130^{+0.177}_{-0.177}}  & \stackval{11.181^{+0.165}_{-0.165}}{11.247^{+0.168}_{-0.168}}  & \stackval{11.099^{+0.182}_{-0.182}}{11.175^{+0.163}_{-0.163}}  & \stackval{11.144^{+0.621}_{-0.621}}{11.257^{+0.264}_{-0.264}}  & \stackval{11.053^{+0.187}_{-0.187}}{11.138^{+0.168}_{-0.168}}  & \stackval{10.986^{+0.222}_{-0.222}}{11.061^{+0.255}_{-0.255}}  \\
        & $\alpha$  & \stackval{-0.353^{+0.057}_{-0.057}}{-0.353^{+0.058}_{-0.058}}  & \stackval{-0.320^{+0.121}_{-0.121}}{-0.319^{+0.118}_{-0.118}}  & \stackval{-0.415^{+0.108}_{-0.108}}{-0.425^{+0.094}_{-0.094}}  & \stackval{-0.353^{+0.109}_{-0.109}}{-0.370^{+0.123}_{-0.123}}  & \stackval{-0.374^{+0.240}_{-0.240}}{-0.388^{+0.204}_{-0.204}}  & \stackval{-0.432^{+0.106}_{-0.106}}{-0.453^{+0.103}_{-0.103}}  & \stackval{-0.551^{+0.222}_{-0.222}}{-0.568^{+0.152}_{-0.152}}  \\
        & $\beta$  & \stackval{-2.858^{+0.912}_{-0.912}}{-2.693^{+0.796}_{-0.796}}  & \stackval{-4.083^{+1.512}_{-1.512}}{-3.641^{+1.245}_{-1.245}}  & \stackval{-5.000^{+3.147}_{-3.147}}{-4.438^{+2.167}_{-2.167}}  & \stackval{-5.000^{+5.147}_{-5.147}}{-4.118^{+1.577}_{-1.577}}  & \stackval{-5.000^{+24.484}_{-24.484}}{-5.000^{+5.025}_{-5.025}}  & \stackval{-5.000^{+3.668}_{-3.668}}{-4.081^{+1.475}_{-1.475}}  & \stackval{-5.000^{+2.907}_{-2.907}}{-4.204^{+2.400}_{-2.400}}  \\
        & MAD  & \stackval{0.045}{0.045}  & \stackval{0.059}{0.060}  & \stackval{0.057}{0.053}  & \stackval{0.033}{0.026}  & \stackval{0.065}{0.045}  & \stackval{0.049}{0.040}  & \stackval{0.079}{0.073}  \\
        \hline

        \multirow{6}{*}{SBPL}
        & $\log_{10}\phi^*$  & \stackval{-2.549^{+0.200}_{-0.200}}{-2.548^{+0.186}_{-0.186}}  & \stackval{-2.634^{+0.206}_{-0.206}}{-2.645^{+0.250}_{-0.250}}  & \stackval{-2.825^{+0.232}_{-0.232}}{-2.865^{+0.250}_{-0.250}}  & \stackval{-2.801^{+0.222}_{-0.222}}{-2.846^{+0.308}_{-0.308}}  & \stackval{-2.989^{+0.372}_{-0.372}}{-3.034^{+0.375}_{-0.375}}  & \stackval{-2.992^{+0.232}_{-0.232}}{-3.030^{+0.206}_{-0.206}}  & \stackval{-3.202^{+0.365}_{-0.365}}{-3.237^{+0.262}_{-0.262}}  \\
        & $\log_{10}M^*$  & \stackval{11.044^{+0.537}_{-0.537}}{11.043^{+0.465}_{-0.465}}  & \stackval{11.027^{+0.379}_{-0.379}}{11.095^{+0.445}_{-0.445}}  & \stackval{11.094^{+0.114}_{-0.114}}{11.177^{+0.321}_{-0.321}}  & \stackval{11.064^{+0.110}_{-0.110}}{11.177^{+0.625}_{-0.625}}  & \stackval{11.103^{+0.359}_{-0.359}}{11.229^{+0.294}_{-0.294}}  & \stackval{11.014^{+0.181}_{-0.181}}{11.037^{+0.287}_{-0.287}}  & \stackval{10.886^{+0.355}_{-0.355}}{10.948^{+0.153}_{-0.153}}  \\
        & $\alpha$  & \stackval{-0.356^{+0.059}_{-0.059}}{-0.356^{+0.059}_{-0.059}}  & \stackval{-0.327^{+0.128}_{-0.128}}{-0.322^{+0.129}_{-0.129}}  & \stackval{-0.425^{+0.096}_{-0.096}}{-0.429^{+0.096}_{-0.096}}  & \stackval{-0.367^{+0.110}_{-0.110}}{-0.370^{+0.131}_{-0.131}}  & \stackval{-0.400^{+0.154}_{-0.154}}{-0.399^{+0.155}_{-0.155}}  & \stackval{-0.447^{+0.097}_{-0.097}}{-0.464^{+0.107}_{-0.107}}  & \stackval{-0.555^{+0.207}_{-0.207}}{-0.561^{+0.157}_{-0.157}}  \\
        & $\beta$  & \stackval{-2.529^{+2.325}_{-2.325}}{-2.382^{+1.781}_{-1.781}}  & \stackval{-3.349^{+3.018}_{-3.018}}{-3.399^{+3.079}_{-3.079}}  & \stackval{-3.956^{+1.448}_{-1.448}}{-3.699^{+3.077}_{-3.077}}  & \stackval{-4.965^{+2.774}_{-2.774}}{-4.140^{+5.921}_{-5.921}}  & \stackval{-5.000^{+29.955}_{-29.955}}{-4.824^{+5.302}_{-5.302}}  & \stackval{-4.657^{+3.427}_{-3.427}}{-3.263^{+2.054}_{-2.054}}  & \stackval{-3.914^{+3.269}_{-3.269}}{-3.423^{+0.935}_{-0.935}}  \\
        & $s$  & \stackval{0.317^{+0.711}_{-0.711}}{0.334^{+0.646}_{-0.646}}  & \stackval{0.107^{+1.842}_{-1.842}}{0.254^{+0.737}_{-0.737}}  & \stackval{0.000}{0.166^{+0.482}_{-0.482}}  & \stackval{0.006}{0.269^{+0.719}_{-0.719}}  & \stackval{0.000}{0.005}  & \stackval{0.004}{0.119^{+0.813}_{-0.813}}  & \stackval{0.003}{0.003}  \\
        & MAD  & \stackval{0.044}{0.044}  & \stackval{0.058}{0.060}  & \stackval{0.054}{0.051}  & \stackval{0.026}{0.026}  & \stackval{0.055}{0.036}  & \stackval{0.043}{0.037}  & \stackval{0.072}{0.065}  \\
        \hline
\end{tabular}
\end{table}
\clearpage

\begin{multicols}{2}
        \section{KiDSDR4-C0v1.1 Catalog Summary}
        \label{app:c0v11_catalog}
        This supplementary section summarizes the content of the intended KiDSDR4-C0v1.1 catalog. The detailed sample selection and filtering procedures are described in the main text. The catalog is provided as a FITS binary table. Table~\ref{tab:c0v11_columns} lists all columns included in the catalog.
        \end{multicols}
    \begin{table}[H]
        \caption{Columns in the KiDSDR4-C0v1.1 catalog.}
        \label{tab:c0v11_columns}
        \begin{tabular}{lll}
            \hline
            \hline
            Column name                 & Unit & Description                                                                                                   \\
            \hline
            \texttt{ID}                 & --   & Unique source ID in the KiDS merged catalog.                                                                  \\
            \texttt{GaZNet-z}           & --   & Deep-learning photometric redshift estimate.                                                                  \\
            \texttt{MAG\_GAAP\_r}       & mag  & KiDS $r$-band GAAP magnitude.                                                                                 \\
            \texttt{MAG\_AUTO}          & mag  & SExtractor AUTO magnitude.                                                                                                 \\
            \texttt{$\log (M_*/M_\odot)$\_CICB19} & dex  & Aperture-corrected $\log(M_*/M_\odot)$ from CIGALE CB19.                                             \\
            \texttt{$\log (M_*/M_\odot)$\_CIBC03} & dex  & Aperture-corrected $\log(M_*/M_\odot)$ from CIGALE BC03.                                             \\
            \texttt{$\log (M_*/M_\odot)$\_CIM05}  & dex  & Aperture-corrected $\log(M_*/M_\odot)$ from CIGALE M05, with an additional $-0.05$ dex offset.       \\
            \texttt{$\log (M_*/M_\odot)$\_LPBC03} & dex  & Aperture-corrected $\log(M_*/M_\odot)$ from LePhare BC03.                                            \\
            \texttt{$\log (M_*/M_\odot)$\_LPM05}  & dex  & Aperture-corrected $\log(M_*/M_\odot)$ from LePhare M05\_tau, with an additional $-0.05$ dex offset. \\
            \hline
        \end{tabular}
    \end{table}

\bibliographystyle{Configurations/scpma-custom}
\bibliography{Configurations/Brief_KiDSDR4SMF}